\documentclass[sigconf, nonacm, 10pt]{acmart}

\AtBeginDocument{%
  \providecommand\BibTeX{{%
    \normalfont B\kern-0.5em{\scshape i\kern-0.25em b}\kern-0.8em\TeX}}}

\usepackage{graphicx}
\usepackage{subcaption}
\usepackage{enumitem}
\usepackage{balance}
\usepackage{xspace}
\usepackage{pifont}
\newcommand{\cmark}{\ding{51}}%
\newcommand{\xmark}{\ding{55}}%
\usepackage[shortcuts]{extdash}
\definecolor{todo-color}{rgb}{1,0,0}
\definecolor{matt-color}{HTML}{228B22}

\newcommand{\tinyskip}{\vspace{3pt}}
\newcommand{\mypar}[1]{\tinyskip\noindent\textbf{#1.}\xspace}

\usepackage{xcolor}
\usepackage{soul}
\usepackage{marginnote}

\sethlcolor{yellow}
\newcommand{\update}[3][0em]{#3}
\newcommand{\updateL}[3][0em]{#3}

\begin{document}


\title{Starling: A Scalable Query Engine on Cloud Function Services}

\author{Matthew Perron}
\affiliation{MIT CSAIL}
\email{mperron@csail.mit.edu}
\author{Raul Castro Fernandez}
\affiliation{MIT CSAIL}
\email{raulcf@csail.mit.edu}
\author{David DeWitt}
\affiliation{MIT CSAIL}
\email{david.dewitt@outlook.com}
\author{Samuel Madden}
\affiliation{MIT CSAIL}
\email{madden@csail.mit.edu}

\begin{abstract} 
  Much like on-premises systems, the natural choice for running database analytics workloads in the cloud is to provision a cluster of nodes to run a database instance. However, analytics workloads are often bursty or low volume, leaving clusters idle much of the time, meaning customers pay for compute resources even when unused. The ability of cloud function services, such as AWS Lambda or Azure Functions, to run small, fine granularity tasks make them appear to be a natural choice for query processing in such settings. But implementing an analytics system on cloud functions comes with its own set of challenges. These include managing hundreds of tiny stateless resource-constrained workers, handling stragglers, and shuffling data through opaque cloud services. In this paper we present Starling, a query execution engine built on cloud function services that employs number of techniques to mitigate these challenges, providing interactive query latency at a lower total cost than provisioned systems with low-to-moderate utilization. In particular, on a 1TB TPC-H dataset in cloud storage, Starling is less expensive than the best provisioned systems for workloads when queries arrive 1 minute apart or more. Starling also has lower latency than competing systems reading from cloud object stores and can scale to larger datasets.
\end{abstract}

\maketitle

\section{Introduction}
\label{sec:introduction}

Modern organizations are increasingly turning to cloud providers to run their data services, including
their database analytics workloads, as witnessed by the proliferation of cloud database analytics products,
such as Amazon Redshift and Microsoft Azure SQL Data Warehouse. \update{R9D2}{These cloud systems avoid the up-front cost of on premises solutions, and allow users to be significantly more elastic.}  However, these systems still typically require users to provision
a cluster of compute nodes of a particular size to run queries.
Unfortunately, because many analytics workloads are unpredictable
and ad-hoc, provisioning is difficult, and often results in over-provisioning, where resources
are underutilized much of the time.  
  Although some cloud services provide ``elastic'' features that
allow compute nodes to be added or removed dynamically, this scaling often takes minutes, making it impractical 
on a per query basis.  Further, 
many cloud database systems require data to be explicitly loaded into proprietary formats. \update{R5W2, R5D6}{For workloads that touch data a limited number of times, such as one-off queries or ETL queries, loading data results in an increase in query latency. Furthermore, cloud storage tends to be an order of magnitude cheaper than other storage services. As a result, several systems, including Presto~{\cite{presto}} and Athena~{\cite{athena}},  are purpose built for executing queries directly on cloud storage. Other systems like Redshift~{\cite{redshift}} have special mechanisms for reading from cloud storage.}

In contrast to current offerings, an ideal system would avoid pre-provisioning servers 
to process or store their data, charge users query-by-query, and be performance competitive. It would also not require users to load data and let users tune to their cost and performance needs on a query-by-query basis.
Although achieving all of these goals perfectly and simultaneously is not possible, 
so called ``serverless'' cloud function services, like AWS Lambda~\cite{lambda} and Azure
Functions~\cite{azure-functions}, offer a tantalizing promise that suggests they may
be able to get close.
In particular, these services allow arbitrary numbers of  small tasks to be invoked with  
very low startup times (typically a few milliseconds) and
offer virtually
unlimited parallelism.  In addition, users
are charged only for the execution time used, typically
at the granularity of 1 second or less.  Using such tasks,
one could invoke many small parallel jobs to scan, join, and aggregate tables in raw cloud storage
using  well known techniques from parallel databases to 
implement a SQL query processing system. 

However, using function services to support ad-hoc analytical
workloads comes with its own set of hurdles.
First, workers or functions have limited memory, execution
time limits, and networking restrictions that prevent  sending data directly
from one instance to another. In  addition, instances are typically
stateless, which is
at odds with stateful analytical queries that need to shuffle or aggregate data.
Thus, in order to support shuffles, function services require other methods of moving data between instances.
Finally, latency of individual worker can be unpredictable, leading  to {\it stragglers} taking much longer
to run than other workers on data of similar scale;  this is particularly true when workers
use proprietary, closed-source, and otherwise
opaque cloud storage services to exchange state, as these services often yield variable and unpredictable
latencies.

To explore the promise of function services for database analytics, 
we built Starling, a query execution engine that runs on serverless platforms. Starling leverages the benefits of cloud services while mitigating the above challenges. To achieve high resource utilization, Starling
maps tasks to function invocations so users pay for only the compute resources
their query actually uses. The number of invocations can grow and shrink as
needed during each query execution. Starling takes advantage of the on-demand
elasticity of cloud object storage services, such as Amazon S3\cite{s3} to
shuffle data. It materializes intermediate results in a format optimized to
reduce cost in the pay-by-request model of these services while achieving high
aggregate throughput. To mitigate stragglers, Starling
uses a tuned model to detect straggling requests and takes
steps to mitigate their impact on query latency. Finally, Starling provides
opportunities to optimize queries for cost or latency by adjusting the number of
invocations at each stage. The ability
to tune queries to cost or performance is a desirable feature for users who run ad-hoc workloads.

With these optimizations, Starling achieves query latency comparable to
provisioned systems while decreasing cost
for workloads with moderate query volume.  

We begin by exploring the properties of available tools for
analytical workloads, and show that Starling fits a point in the
design space that has so far remained unaddressed.

\section{Motivation and Design}
\label{sec:design}

\begin{figure*}
  \centering
  \includegraphics[width=\textwidth]{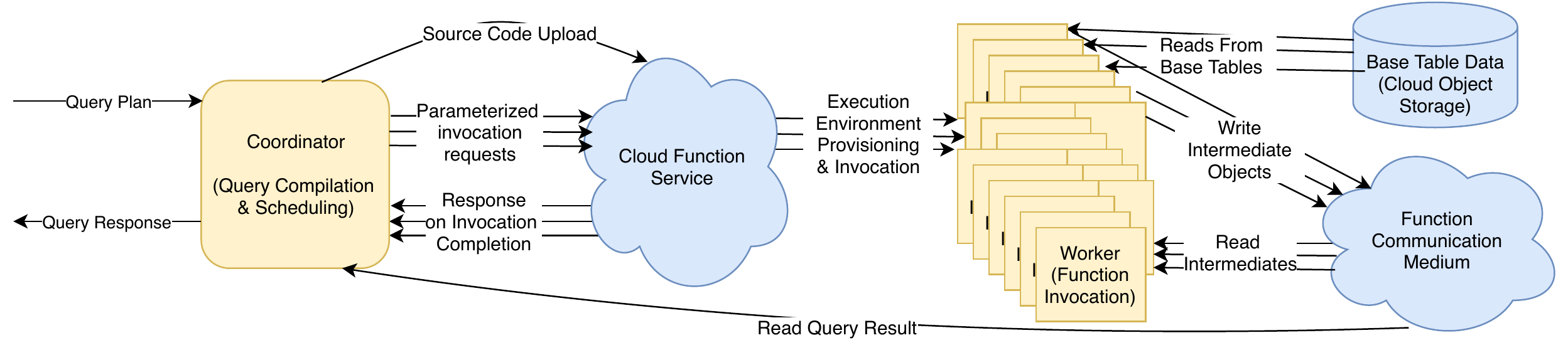}
  \caption{Query Execution in Starling. Opaque cloud components in blue, Starling components in yellow}
  \label{fig:system_diagram}
\end{figure*}

Starling seeks to provide a balance of performance, flexibility, and low cost-per-query that current systems
do not provide for some important classes of workload. Below we describe the current landscape of systems and describe the promise and challenges
of cloud functions for query processing. We follow with a brief description of the architecture of
Starling, and conclude by discussing why we chose Amazon AWS to for our Starling implementation.

\subsection{Landscape of Cloud Analytic Databases}
\label{subsec:landscape}

\begin{table}
\small
\begin{tabular}{|c|c|c|c|}
\hline
System & Does not & Pay by & Tunable \\
& require loading& query  & performance \\
\hline
Amazon Athena & \cmark & \cmark & \xmark \\
Snowflake & \xmark & \cmark* & \cmark \\
Presto & \cmark & \xmark & \cmark \\
Amazon Redshift & \xmark & \xmark & \cmark \\
Redshift Spectrum & \cmark & \xmark & \cmark \\
Google BigQuery & \cmark & \cmark & \xmark \\
Azure SQL DW & \cmark & \xmark & \cmark \\
\textbf{Starling} & \textbf{\cmark} & \textbf{\cmark} & \textbf{\cmark} \\
\hline

\end{tabular}
\caption{Comparison of cloud analytics databases}
\label{table:landscape}
\end{table}

The proliferation of cloud analytic databases has led to a rich ecosystem of
offerings with varying features and pricing models. Table~\ref{table:landscape} shows an overview of the
design space. The rows in the table correspond to some of the most popular analytic databases. The columns are as follows:

$\bullet$~\textbf{Does not require loading} Some systems need to load data from its
  original format, e.g. CSV, ORC, Parquet, etc., to an internal format that permits 
executing queries with high performance. This loading step is a barrier
to users who want to run ad-hoc queries on raw files in cheap cloud object storage.
With the rise of enterprise data lakes, having analytic data stored in raw formats in inexpensive cloud object stores is increasingly the norm. \updateL{R9D3}{While many systems have methods of reading data from external sources like cloud storage, this is typically a tacked-on option that results in significant performance degradation compared to data stored on local disks in native formats.} 

$\bullet$~\textbf{Pay per Query} Provisioned systems start a cluster that sits in the cloud
waiting for queries to process. Whether the system is idle or not, the cloud
vendor charges for the underlying virtual machines (plus some fixed cost for the
data analytics service). An alternative, in a  \emph{pay-per-query} model, users are
only charged based on the queries they run. Such a model can be dramatically
cheaper if queries are issued sporadically or unpredictably. We evaluate specifically
 when pay-per-query is
more cost-efficient than Starling in our experiments. \updateL{R9D3}{While Snowflake does not
have a pay-per-query model per se (hence the asterisk in the table), it allows users to automatically shut down clusters during periods of inactivity, and resume processing when new queries arrive, saving users money when query volumes are low.}

$\bullet$~\textbf{Tunable Performance} In cloud settings, both the response time and
cost of executing a query depends on the amount of resources that are
provisioned. When systems permit scaling  resources to tune
performance and cost in response to data volumes, then we say these systems are elastic.
When systems are inelastic or do not allow tuning, queries may fail to execute
or take longer than users require. \updateL{R9D3}{The range of elasticity varies between systems varies. While
Redshift configurations may take minutes to add new nodes, Snowflake allows users to start ``virtual warehouses'' of varying size with latency of a few seconds.}

Starling is a system for analytic users who run a low to moderate query volumes 
on data in cloud object storage. 
It i) does not
require loading; ii) charges only for the queries that are executed; and iii) permits users to
trade off cost and performance, and adjust parallelism on a per-query basis.

None of the existing cloud systems offer all three options, as shown in
Table~\ref{table:landscape}. Cloud functions are the building block that allows us to 
achieve these objectives.

\subsection{Cloud Functions}
\label{subsec:cloud_functions}

Cloud functions services, also known as Functions as a Service (FaaS) allow
users to run applications without managing or provisioning servers. Users upload
application code or executables to the service. In response to events or direct user
invocation, the function service provisions an execution environment
 and runs the user-provided code. For our purposes, the key advantage of cloud functions are  i) they can read directly from cloud storage,  ii)
 they have a very low startup time and are billed on a per-invocation basis
 and iii) many of them can be invoked in parallel. These properties translate directly to the desired features
 in Starling (no loading, pay-per-query, and tunable parallelism/performance.)

Unfortunately, despite their high-level appeal,
analytical workloads are not a natural fit for
cloud functions for several reasons. First, an analytic query can run for
hours, but cloud function execution is limited to a few
minutes. Moreover, cloud functions execute in resource-constrained containers; e.g., 
  1 core per function and at most 3 GB RAM is typical of current implementations.
Second, analytic queries require shuffling data to compute joins, but cloud
functions do not allow communication between invocations to enforce isolation and security
policies. Although these limitations have been identified as show-stoppers
before~\cite{hellerstein2018serverless}, in this paper we develop mechanisms to work around these
shortcomings and deliver a performance and price-competitive data analytics system built
on cloud functions.

\subsection{Starling Architecture}

Starling is a query execution engine. Users submit planned queries to the system
and receive back query results. Figure~\ref{fig:system_diagram} shows Starling's
architecture. As seen in the figure, users submit queries to a small Coordinator
that compiles the query and uploads it to a cloud function service.
The coordinator then schedules the tasks by invokes them through the function service. The function
service is responsible for provisioning execution environments for Workers that perform
the task of query execution. Workers read base table data from inexpensive cloud
object storage services. Because functions are stateless they exchange state through
a communication medium, e.g., shared storage. When all tasks complete the worker reads the result
from the function communication medium and returns it to the user. 

\update{R6W2}{Starling's design requires a few properties of underlying cloud services. First, Starling needs to launch hundreds of function invocations at once and in parallel from cold start. Second, Starling relies on relatively inexpensive and high throughput methods of exchanging data between tasks, such as object storage services.  Its performance depends on individual parallel workers being able to each achieve high throughput despite other workers executing concurrently.}

Next, we describe the division of work between the Coordinator and Workers in more detail.

\mypar{Coordinator}
The coordinator compiles queries, uploads the executable to the cloud function
services, and manages the execution of the query. It can run on a small virtual
machine. Starling has no query optimizer, so the coordinator takes planned queries
as input, and compiles the query into an executable that can execute any task of
the query plan. We describe this process in Section~\ref{sec:qe}. It uploads the
executable to the function services, packaged with necessary supporting files.
Starling then schedules tasks to execute the query end-to-end,
 assigning tasks to workers and monitoring their completion. 
 Decomposing queries into tasks, and determining which tasks
 to run when is the job of Starling's scheduler.

\mypar{Workers}
Each worker runs a single cloud function invocation to execute part of a query. A
worker is invoked with parameters indicating
the task to be executed. These parameters are the only communication from the
coordinator during the lifetime of the worker. Because function invocations
cannot communicate directly, they must use some communication medium, such as
shared storage,
to exchange data.
The worker reads inputs from either base table storage or the
communication medium, processes them and writes its output to the communication
medium. For operations like parallel joins or aggregations that need
to shuffle data, workers that produce intermediate output write individual partitions.
Consuming workers read these partitions. A detailed description of how Starling
manages this data is given in Section~\ref{sec:storage}, while details of 
how workers execute
queries efficiently are given in Section~\ref{sec:qe}. After the function writes its
outputs, the worker exits. The function service notifies the coordinator that
the worker has completed, and query execution continues. 

\subsection{Choosing a Cloud Function Service}
\label{subsec:amazon_services}

Google Cloud Functions~\cite{google-functions}, Azure Functions~\cite{azure-functions},
 and AWS Lambda~\cite{lambda},
each provides a function service. Starling's architecture can be implemented in
any of these platforms, however, certain platforms offer features that are
more amenable to the kind of workloads Starling is designed to support. In particular,
Google Cloud Functions and Azure functions have restrictions on the
languages that can be used, or the rate at which functions can be invoked,
or both, that will impact parallel query
processing performance, while AWS Lambda does not have such restrictions.

For these reasons we chose to build Starling on AWS Lambda.
This restricted us to using AWS services to exchange
intermediate data.
In the next section we discuss
why we chose S3~\cite{s3}, AWS's cloud object storage offering, for both
base table storage and as the communication medium.

\section{Managing Data in Starling}
\label{sec:storage}

Storage is an important component of any data management system. Starling provides interactive query performance on raw data stored in cloud object storage. Starling does not manage base table data, but must interact with it efficiently, as we describe below in Section~\ref{subsec:storage_base_tables}. Because cloud functions are stateless, Starling must also manage intermediate state during query execution; 
we describe this process in detail in Section~\ref{subsec:storage_intermediates}. 

\subsection{Base Table Storage}
\label{subsec:storage_base_tables}

Starling executes queries over data sitting in S3. Starling's design is agnostic to base table formats, but common choices are CSV, ORC, and Parquet. Starling requires only that rows of a specified schema can be parsed from the source objects in S3; however, for the best performance base table data should be stored as objects of a few hundred MB.

Open source columnar formats like ORC~\cite{orc} help Starling to achieve good
performance as they allow reading a subset of columns rather than the whole row.
This helps workers in Starling save time by skipping columns that are not
needed. ORC also includes indexes and basic statistics that allow users to skip
portions of the input increasing performance. Therefore, in our evaluation in Section~\ref{sec:evaluation} we use raw data stored in ORC.

\subsection{Managing Intermediate State}
\label{subsec:storage_intermediates}
\begin{figure}
  \centering
  \includegraphics[width=0.95\columnwidth]{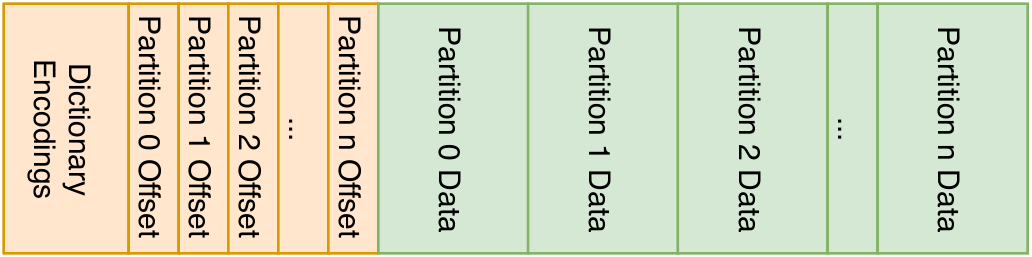}
  \caption{Starling Partitioned S3 Object Format. Metadata in orange, partitioned data in green}
  \label{fig:file_format}
\end{figure}

As cloud functions are stateless and have no method of communicating directly, Starling relies on AWS services to shuffle data. A medium for exchanging data between function invocations should have low cost, high throughput, low latency, and scale transparently. We considered several options for exchanging intermediate state before choosing Amazon's cloud object storage service, S3. Using virtual machines or a streaming system like Amazon Kinesis~\cite{kinesis} both require users to provision capacity ahead of time and thus are not a suitable choice. We also considered using Queue services like Amazon SQS~\cite{sqs}, but these limit message sizes (to 256KB in the cases of SQS) and, in the case of SQS require encoding data as text, making it cumbersome and computationally costly for large shuffles. NoSQL services like DynamoDB~\cite{dynamo} have very low latency but unacceptably high cost for large shuffles. While S3 has relatively high latency compared to some of these alternatives, this can be mitigated as we describe in Section~\ref{subsec:storage_latency}. We describe the most important properties of S3 for Starling below.

\textbf{S3 Properties:} S3~\cite{s3} is an AWS's object storage service. Users write binary objects of arbitrary size to the service into ``buckets'' with a named ``key''. S3 is a write-once system, meaning that no appends or updates to objects are allowed, only replacement. Users can scale read and write throughput by ensuring keys are spread across ``prefixes'', defined as the first few characters in the key~\cite{s3-performance}. Users interact with the service through a REST interface. Reads from S3 can fetch the entire object or a range of bytes. S3 charges users by the amount of data stored at a rate \$0.23 per GB per month, and a cost of \$0.0004 per thousand GET requests of any size, and \$0.005 per thousand \texttt{PUT}s (prices as of July 2019). Unlike standard file systems, S3 does not guarantee read-after-write consistency, which, in some cases complicates query processing as we describe in Section~\ref{subsec:visibility_stragglers}. S3 provides atomic reads and writes, ensuring that readers never see data from two separate writes in the same read.

\textbf{Sharing intermediates:} Starling uses S3 to pass intermediate data between function invocations. Workers
write their outputs as a single object in S3 with a predetermined key. Because
the object is written at a known location, readers can poll the object key until
the object appears. For query processing, S3 has the additional advantage of
being persistent, meaning workers can begin sending data before destination
workers have started executing. This saves a significant amount of execution
time in AWS Lambda.

Using S3, one to many communication is both inexpensive and straightforward.
Producer tasks write an object to S3, making it visible to all readers that need
it. All-to-all communication, as in a shuffle, is more difficult to
achieve at low cost. As recent work has demonstrated, writing an object per
partition for large shuffle incurs unacceptably high
cost~\cite{pu2019shuffling}. Starling ameliorates this problem by writing a
single partitioned file to S3 and having consumer tasks read only the relevant
portion of each object output by the producers. We describe this process in more
detail in Section~\ref{subsec:shuffling}.

Producer tasks in Starling each write a single object to S3 containing all
partitions, to avoid excessive write costs. Figure~\ref{fig:file_format} shows
the format of these partitioned files. Each file contains metadata containing
the end location of each partition in the object. To decrease the latency of
reading and writing objects to S3, Starling encodes low cardinality string
columns using dictionary encoding~\cite{westmann2000compressed}. The dictionary encoding is included
at the beginning of the file. Metadata is followed by the partitioned data.
Optionally, partitions may be compressed with general-purpose compression.

To read the intermediate file, consumers make one read to fetch  metadata at the
head of the object, then make a subsequent read to fetch the needed partition.
This allows consumers to read any partition with two reads. In addition, this
format also allows consumers to easily fetch adjacent partition data with the same
number of \texttt{GET} requests as a single partition. We use this to support
multi-stage shuffles as discussed in Section~\ref{subsec:shuffling}

\subsection{Mitigating High Storage Latency}
\label{subsec:storage_latency}

\begin{figure}[t]
  \centering
  \includegraphics[width=0.95\columnwidth, trim={0 0.25cm 0 0.25cm},clip]{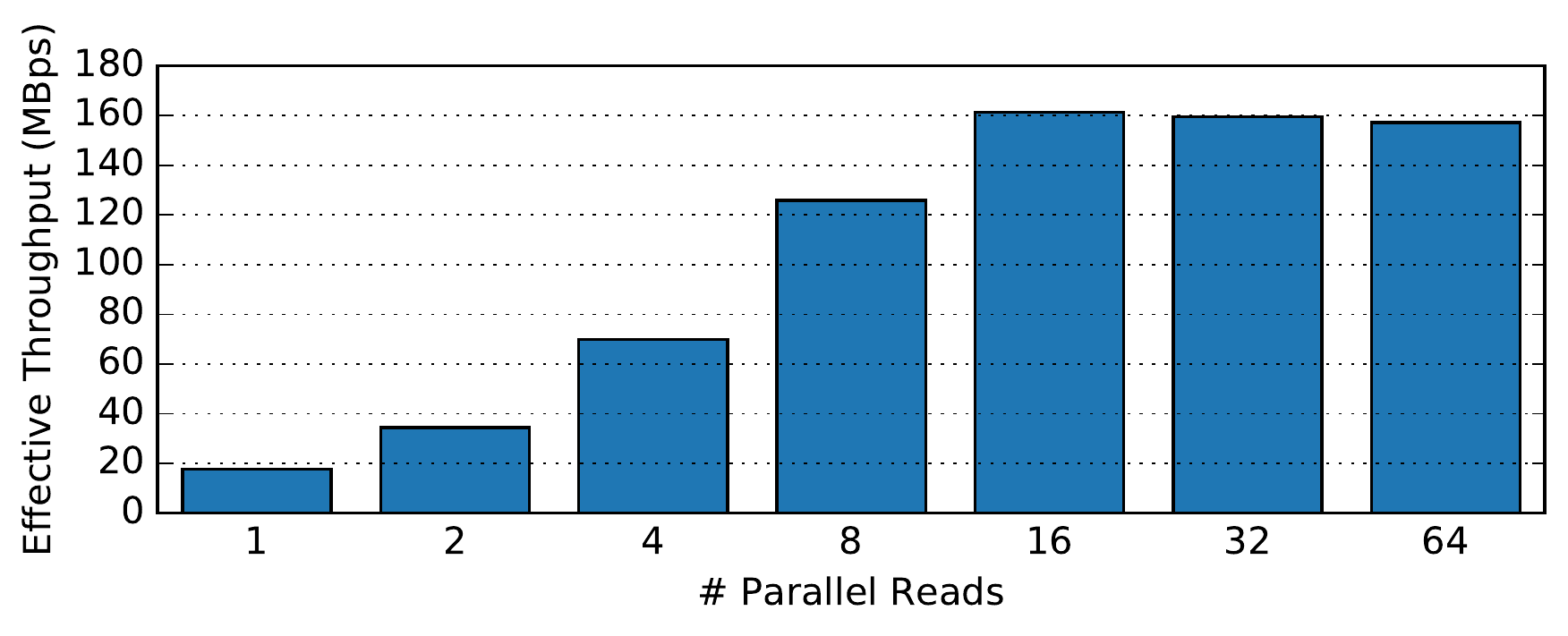}
  \caption{Effective throughput of a single function invocation with increasing parallel 256KB reads}
  \label{fig:parallel_reads}
\end{figure}

While S3 has very high aggregate throughput, it has much higher latency than
other shuffling options. A 256KB read has median latency of 14ms. If workers
perform single-threaded blocking reads to the S3, they may sit idle while
waiting for responses, increasing query latency and thus increasing invocation
runtime costs. In order to mitigate this latency each task performs several
reads in parallel from S3. Fortunately, most tasks in Starling must perform many
reads, making this trivial to parallelize. Columnar formats like ORC are broken into column segments, and in joins, tasks must make many reads to fetch partition data from objects written by input tasks. Parallelizing reads helps keep tasks busy, spending less time idle and more time on query processing. Figure~\ref{fig:parallel_reads} shows the total throughput of a function invocation reading many 256KB reads. As we increase the number of parallel reads that the invocation performs, we achieve higher effective throughput until 16 parallel reads, after which adding more parallel reads does not improve latency. 

\subsubsection{Mitigating Object Visibility Latency}
\label{subsec:visibility_stragglers}

As noted above S3 does not guarantee read-after-write consistency in some cases. As a result, an object recently written to S3 may not become visible to tasks in a subsequent stage for several seconds or more. Although this is infrequent, because a shuffle is an all-to-all communication, any delay in object visibility slows down all reading workers and can have detrimental impact on query latency. Furthermore, tasks reading the object continue to incur delays waiting for it to become readable. Starling mitigates this risk by writing the same object to two different keys in S3. We call this optimization ``doublewrite''. Consumers attempt to read the first key, and if it is not available, try the second one. This strategy makes query performance more predictable by reducing the risk that a single visibility issue slows down all consumers.

\section{Query Execution}
\label{sec:qe}

The goal of Starling's query execution engine is to achieve interactive performance at low cost. To run queries on AWS Lambda, Starling's coordinator uses a single JSON file describing a physical query plan as input. \update{R6W1, R6D1}{The plan contains dependencies between stages, as well as the number of tasks within each stage. The coordinator monitors task completion and starts new stages once dependencies are completed. More details follow in Sections~{\ref{subsec:tasks}} and {\ref{subsec:pipelining}}}. The coordinator generates C++ source code for the query, and compiles it into a single executable that is then packaged with necessary dependencies, compressed in an archive and uploaded to AWS Lambda.  \update{R6W1, R6D1}{Each task's input and output object names are determined before each stage begins execution.} Each task executes as much of the query as possible without communicating with other workers. For instance, Starling's workers may perform multiple joins if tables are partitioned on the same key. The coordinator then invokes query tasks until query completion. Below we describe how relational operators are implemented, how shuffles are performed, and how tasks are scheduled to balance performance and cost.

\begin{figure}
  \centering
  \begin{subfigure}[b]{\columnwidth}
    \includegraphics[width=0.95\columnwidth,  trim={0 1cm 0 1cm},clip]{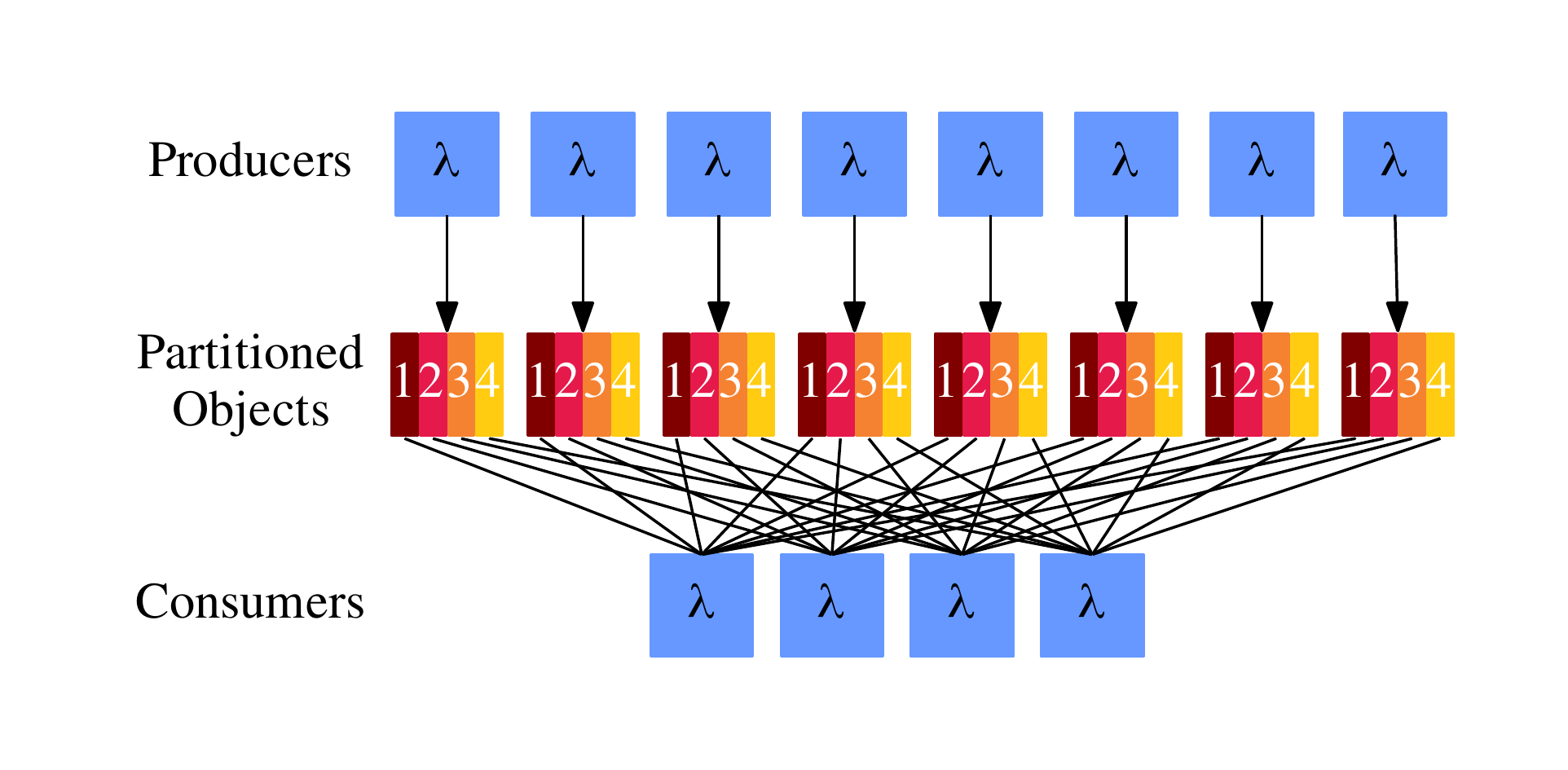}
    \subcaption{Standard Shuffle}
    \label{fig:std_shuffle}
\end{subfigure}
\begin{subfigure}[b]{\columnwidth}
  \includegraphics[width=0.95\columnwidth, trim={0 1.2cm 0 1cm},clip]{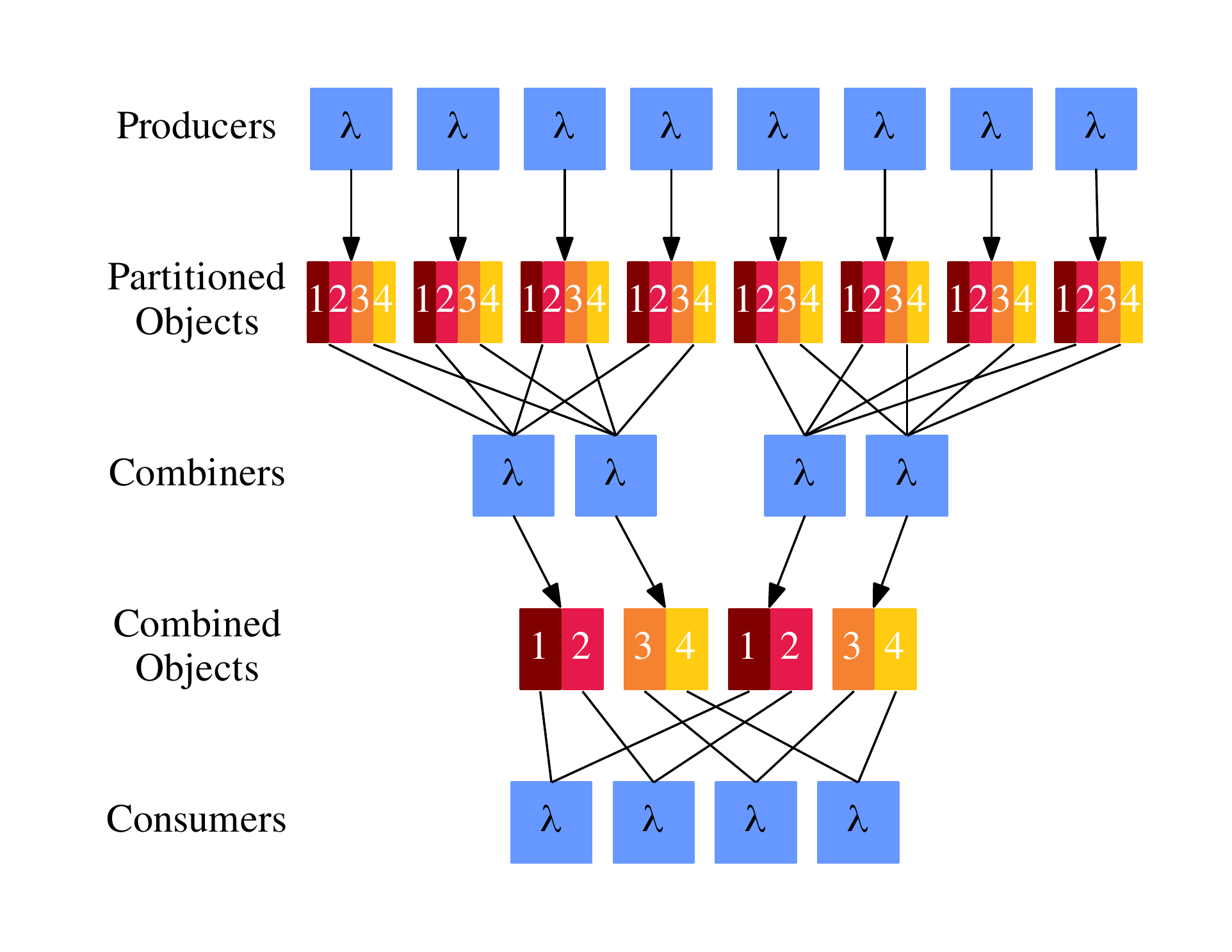}
    \subcaption{Multistage Shuffle}
    \label{fig:ms_shuffle}
\end{subfigure}
\caption{\update{R6NP}{Starling shuffling strategies, function executions in blue, S3 Objects in shades of red showing partitions. Lines are reads and arrows are writes}}
\end{figure}

\subsection{Relational Operator Implementation}
\label{subsec:worker_implementation}

After reading data from S3, workers in Starling execute using data-centric
operations, where operators are implemented as a series of nested loops, rather than a pull-based approach. Query compilation allows for type
specialization and achieves very good performance for analytics use cases
systems~\cite{kersten2018everything}. Essentially each task contains a set of
nested loops each performing necessary relational operations. This pipelining of
operations within workers contributes to Starling's low query latency. The tasks
write the materialized output of its operations as a single object to S3. 

Below we describe how relational operators are implemented in Starling.

\mypar{Base Table Scans}
While Starling does not manage base table data directly, it still must be able
to scan data from base tables quickly. It does this by reading portions of input
files in parallel. If there is a projection, Starling reads only the necessary
columns from the base table, if possible given the file format (e.g., in ORC or Parquet).

\mypar{Joins}
Starling supports both broadcast joins and partitioned hash joins. In the case
of broadcast joins, each input task for the inner relation writes a single
object to S3. On the outer relation, tasks read all data from inner relation and
their own subset of the outer relation to perform the join.

Partitioned hash joins require a shuffle. Tasks scan both relations and
partition their data on the join key and write partitioned files as described in
Section~\ref{subsec:storage_intermediates}. If possible, this partitioning is
pipelined in a single task with other operations. Afterwards, a set of join tasks
is started to perform the join. These join tasks create a hash table for their
partition of one relation and then scan the other relation, probing into the
hash table. As shuffling efficiently is critical for performing partitioned
joins, we describe in detail how shuffles are performed in
Section~\ref{subsec:shuffling}.

\mypar{Aggregation}
Starling performs aggregation in two steps. Tasks that perform the final operation before aggregation
each generate a set
of partial aggregates and output an object to S3. To complete the query, a final
task reduces these partial aggregates into a final aggregate. When necessary,
Starling first performs a shuffle, partitioning on the \texttt{group by} key before
generating partial aggregates. 

\subsection{Shuffling}
\label{subsec:shuffling}

As we describe in Section~\ref{subsec:storage_intermediates}, Starling uses a partitioned intermediate format allowing tasks performing a shuffle to only read relevant partitions from input files. 
In a standard shuffle, each consumer must read from every output, an all-to-all communication. Since the workers in Starling are small, this results in each task reading a large number of small objects from the storage service. In addition to impacting query latency, having so many reads may incur an unacceptably high cost since object storage services charge by request. For example, with 512 producer tasks and 128 consumer tasks, the S3 cost for this shuffle is only 5.7 cents at current S3 pricing, but for a larger shuffle of 5120 producers and 1280 consumers, the cost increases to more than \$5. 

To address this issue, we implement two strategies for partitioned hash joins. First, for small joins we implement a standard shuffle as described above where every consumer reads output from every producer task. A diagram of this strategy is shown in Figure~\ref{fig:std_shuffle}. The number of reads is $2sr$ where $s$ is the number of producer tasks and $r$ is the number of consumer tasks. 

Since request costs become unacceptably high for joins with many input tasks, we trade off compute time for object storage request costs by implementing a multi-stage shuffle. We introduce a stage of combining tasks between the producers and consumers. Each task in the combining stage reads a contiguous subset of partitions from a subset of input objects and produces a single combined output, with the same partitioned file format. Because these combining tasks read contiguous partitions they still perform only two reads per input. Finally, the consumer tasks read outputs written by relevant combiners. Because each combining task reads a subset of partitions, read outputs need only read a subset of the outputs of these combining tasks. A diagram of this approach is shown in Figure~\ref{fig:ms_shuffle}. The number of S3 reads in a multistage shuffle is $2(s/p+r/f)$ where $p$ is the fraction of partitions each combiner reads and  $f$ is the fraction of files each combiner reads. The number of combining tasks is $1/(pf)$. Figure~\ref{fig:ms_shuffle} shows a multi stage shuffle with $p=1/2$ and $f=1/2$, where each combining task reads half of the partitions from half of the input files. With 5120 and 1280 consumer tasks and $p=1/20$ and $f=1/64$, the S3 read cost is just \$0.073 compared to more than \$5 with a standard shuffle. The additional write cost of these combiners is negligible. Each of the 1280 combiners makes two additional writes, costing an additional \$0.00128.

In a multistage shuffle, we can create any number of combining tasks, but we typically choose the same number of combining tasks as receiving tasks.

\subsection{Assigning Tasks to Workers}
\label{subsec:tasks}
The primary way that cost and performance is managed in Starling is by controlling the number of tasks per stage. Typically having more tasks per stage results in lower latency but higher cost because of the overhead of exchanging intermediate state between workers. However, this is a delicate balancing act. With too many workers these overheads may overwhelm any potential performance gains. But with too few tasks, resource constrained workers may run out of memory. The space in between these two extremes allows users of Starling to trade off cost and performance by tuning the number of workers per stage. 

For large queries, we sometimes need to execute more tasks than the maximum available parallel function invocations, as Amazon imposes a limit on the total tasks that can be executed at once. In July 2019 the limit on parallel AWS Lambda function invocations is 1,000. This can be increased by contacting Amazon. For our experiment we set maximum number of parallel invocations to 5,000. The coordinator sets a maximum limit on the number of parallel tasks per stage, Once the limit is reached, Starling waits for a task to finish before scheduling a new task, until all stages are complete.

Because Starling does not currently have a query optimizer, we expose user-configured parameters necessary to tune for cost and performance, including the shuffling strategy and the number of tasks per stage.

\subsection{Pipelining}
\label{subsec:pipelining}
Instead of waiting for all tasks in a stage to be complete before starting consuming stages, a strategy for decreasing the latency of queries is to start consuming stages when a large fraction of producer inputs are available. This allows workers to start reading the available inputs and decreases overall query latency, by mitigating the impact of some stragglers. However, this comes with additional risks and costs. If a task in the producer stage straggles during a standard shuffle, it causes all reading tasks to sit idle, significantly increasing the cost of executing this stage. Thus, turning on pipelining typically results in lower query latencies, but comes with additional cost. Users who desire the least expensive query execution possible should disable pipelining. One way to reducing this cost is by eliminating as many stragglers as possible mitigation techniques.

\section{Stragglers}
\label{sec:straggler_handling}

Starling relies on S3 for both reading base table data as well as for exchanging
intermediate state between function invocations. As stages must wait for their
inputs to be available before doing the computational work of query processing,
stragglers in any of these requests can have significant impact on query
latency. Unfortunately, S3 requests often suffer from poor tail latency,
with a small fraction of reads and writes taking much longer to complete. Thus, straggler mitigation is critical in making Starling performance competitive with provisioned systems. One of the chief challenges of mitigating stragglers is that these services are outside the control of Starling, are closed source, and have opaque operation. \update{R5W3, R5D10}{Therefore, we base our optimizations on the power of two choices.{~\cite{mitzenmacher2001power}}, a well known theoretical framework for using randomization and duplicate tasks to improve performance in unpredictable distributed systems, which systems like MapReduce and Spark also employ to good effect.}

\subsection{Read Straggler Mitigation}
\label{subsec:read_straggler_mitigation}

\begin{figure}[t]
  \centering
  \includegraphics[width=0.95\columnwidth, trim={0 .2cm 0 0},clip]{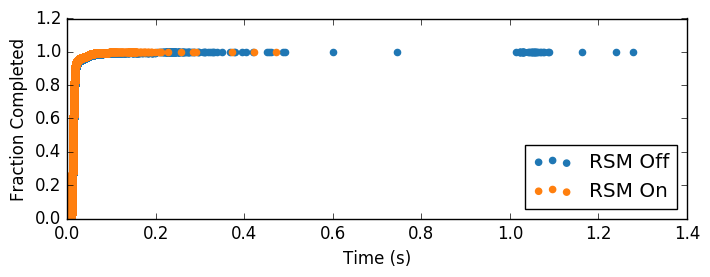}
  \caption{Completion time CDF for 256KB Reads to S3 from AWS Lambda. Comparing read straggler mitigation on and off}
  \label{fig:256K_RSM}
\end{figure}

A single query in Starling may perform
hundreds of thousands of S3 \texttt{GET} requests. Some of these requests experience
significant delays, as we show in Figure~\ref{fig:256K_RSM}. We mitigate these stragglers
by observing how long a request takes compared to its expected completion time.
Workers detect when a request is taking longer than expected, and open a new
connection to S3 and retry the request. Workers use a simple model to
determine when queries should arrive based on observed latency and throughput of
S3 requests as well as the throughput of AWS Lambda Invocations. Starling's model for
expected query response time is given by:
$r = l +
(b/tc)$, where
$r$ is the expected response time, $b$ is the number of bytes requested, and $c$ is
the number of concurrent readers. The tunable parameters of the model $l$ and $t$, correspond
to latency and throughput of AWS Lambda invocations reading from S3. We measure these as 15ms and 150MBps
respectively.

If S3 fails to respond to a request within a fixed factor of the expected time, Starling sends
a duplicate request, accepts whichever response returns first, and closes the
other connection. While we do not have
insight into the design of S3 and thus cannot determine the source of these
stragglers, we find that this strategy mitigates most read stragglers and
significantly improves query latency.

To show this, we evaluate Starling's read mitigation strategy using a microbenchmark, performing thousands of reads to S3 from Lambda. In Figure~\ref{fig:256K_RSM} we show a CDF of the request time for a set of 256KB reads with and without read straggler mitigation (RSM) enabled. While the straggler mitigation mechanism is not perfect and sometimes request takes as long as 2.5 seconds, the long tail of requests seen without this mitigation is cut short, helping queries to run faster. At the 99.99th percentile, latency is more than a second without RSM and .25 seconds with RSM.

Although duplicate requests result in an additional expense, this pays off in saved function invocation time. An additional read request needs to save just 8 milliseconds of invocation time to pay for itself. While the read straggler mitigation mechanism is only triggered in 0.3\% of cases (160 times for 52,000 reads), in this experiment the mitigation saves nearly 95 seconds of compute time at an additional read cost the equivalent of only 1.3 seconds, making read straggler mitigation a cost saving measure in addition to a latency reducing strategy.

\subsection{Write Straggler Mitigation}
\label{subsec:write_straggler_mitigation}

While most queries perform several
orders of magnitude fewer writes to S3 than reads, usually two per function invocation,
write requests tend to be much larger, up to several hundred MB. Median latency
for these large write requests may be several seconds. Write stragglers must
be handled differently. While read requests are small and their responses are
large, the inverse is true for writes. Furthermore, we observed that most
stragglers on write are not due to slow data transmission rates to S3, but in
the S3 service processing requests and sending responses. That is, write
requests are sent to the S3 service quickly, but replies from S3 may be delayed
for unknown reasons. By using a strategy similar to RSM, Starling
may react slowly to cases where data is written to S3 quickly but S3 is sluggish
in its response. Therefore, we use an additional model to predict response times for writes
\textit{once the request has completed sending}. The expected response time is the same
as the RSM model but requires a different set of parameters since the internal throughput of the
S3 service is much higher than the throughput from a single function invocation.
When either of these models for response time indicate that a straggler has occurred
 a second write request is started on a new connection. 

\begin{figure}[t]
  \centering
  \includegraphics[width=0.95\columnwidth, trim={0 .2cm 0 0},clip]{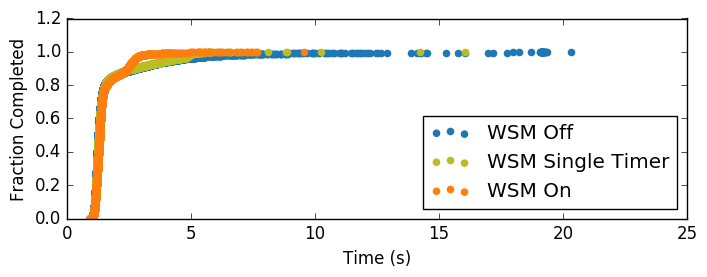}
  \caption{Completion time CDF for 100MB Writes to S3 from AWS Lambda. Comparing write straggler mitigation off, with single timeout only and fully on}
  \label{fig:100M_WSM}
\end{figure}

 We determine how well the write straggler mitigation (WSM) mechanism works by running another microbenchmark. Figure~\ref{fig:100M_WSM} shows the result of running many 100MB writes to S3 and measuring the response time. We compare the response times without write straggler mitigation, with only a single timeout (the same as read straggler mitigation) and with the full write straggler mitigation including a second timeout set after the client finishes sending its request. Without write straggler mitigation, the longest writes take more than 20 seconds. While the single timeout, reduces these worst performing reads to about 18 seconds, the full write straggler mitigation brings the tail latency down to about ten seconds. As writes are much less frequent than reads in Starling, this long tail is encountered much less frequently. At the 99th percentile writes take almost 9 seconds without WSM, 5 seconds with only one timeout, and 3.8 seconds with full write straggler mitigation.

To pay for itself, each additional write would need to save 102 milliseconds of compute time. In this experiment, the full write straggler mitigation is invoked in 31\% of writes, 3138 of 10240 writes, costing the equivalent of 314 seconds of compute, while saving the equivalent of 2100 seconds.

Since all writes have to complete for following stages to read them, WSM is a critical part of Starling achieving low latency while also saving compute time.

\section{Evaluation} 
\label{sec:evaluation}

In our evaluation we seek to answer the following questions in corresponding sections:
\begin{itemize}[leftmargin=*]
  \item How does Starling's operational cost compare to alternatives as query workloads change? (Section~\ref{subsec:eval_cost_comparison})

  \item How performant is Starling compared to alternatives? (Section~\ref{subsec:eval_cost_performance})

  \item How well does Starling scale to larger datasets? (Section~\ref{subsec:scalability})
  \item Can Starling support concurrent queries? (Section~\ref{subsec:concurrency})

  \item How does Starling compare to current cloud interactive pay-by-query services? (Section~\ref{subsec:payg_systems})

  \item How well does Starling allow users to tune for cost and performance? (Section~\ref{subsec:eval_scheduling})

  \item How important are Starling's performance optimizations for achieving low query latency? (Section~\ref{subsec:eval_perf_opt})

\end{itemize}

\subsection{Experimental Setup}
\label{subsec:setup}
We execute our experiments on a scale factor 1,000 (1TB) TPC\=/H~\cite{tpc-h} dataset for most experiments, and scale factor 10,000 (10TB) for the scaling experiment. Each uncompressed table is broken into files of size at most 1GB, then encoded using Apache ORC~\cite{orc}, a standard columnar format. The ORC files use Snappy compression~\cite{snappy} to compress columns. The resulting ORC files are uploaded to a single Amazon S3~\cite{s3} bucket. The size before compression is about 1TB and the size after conversion to ORC is 317GB. This dataset has no skew. \updateL{R5W1, R5D1, R5D2}{We compare to other systems on a subset of TPC-H queries: all queries except 11, 21, and 22. We exclude these queries as Starling is not feature complete, or because a compared system could not execute it. We do not have reason to suspect that the addition of these queries would substantially change our results. For each configuration we execute each query sequentially, and report the median latency of three executions.} We describe each of the systems and configurations that we compare against below.

\mypar{Amazon Redshift}
Amazon Redshift~\cite{redshift} is a data warehouse product sold by Amazon Web Services(AWS). Users provision a cluster with a fixed number of nodes, though nodes can be added later. Users must first pre-load data before it can be queried. Redshift offers two large node types: A ``dense compute'' node type \texttt{dc2.8xlarge} with 32 threads, 244GB of DRAM and 2.56TB of SSD storage, and a ``dense storage'' node type \texttt{ds2.8xlarge} with the same CPU and memory, but with 16TB of HDD storage. Redshift also allows users to read data from external tables in S3 with a feature called Spectrum. Instead of performing reads directly from the user's cluster, Spectrum spawns short lived workers, much like AWS Lambda invocations, to filter base table data from S3 and send it to the cluster to complete the query.

Redshift allows users to divide tables among nodes on a distribution key and sort on a key by defining these keys in the schema. Doing so significantly decreases query latency when tables are distributed on join keys and sorted, as nodes can perform local merge joins without shuffling data. 

We compare against four configurations of Redshift reading from local data, and one reading from S3 using Spectrum. With local data we have two clusters, one with ``Dense compute'' nodes with local SSDs and one with ``Dense Storage'' with 16TB of HDD storage each, abbreviated \texttt{dc} and \texttt{ds} respectively. We also compare two different schemas, one with distribution keys and sorting keys defined, and one without where data is equally partitioned among nodes without respect to their join key and sort order. The configurations on the dense compute cluster with distribution keys we call \texttt{redshift-dc-dk} and with a without distribution keys we call \texttt{redshift-dc-dd}. Likewise the dense storage configurations we call \texttt{redshift-ds-dk} and \texttt{redshift-ds-dd} for distribution keys and default distribution, respectively. 

In addition, we compare with a four node \texttt{dc2.8xlarge} cluster using the Spectrum feature to read all base table from S3. We call this configuration \texttt{spectrum}. 

\update{R5W1, R5D3}{We report the cost of running Redshift's nodes using on-demand pricing {\cite{redshift-pricing}} for all configurations and add S3 scans costs for the {\texttt{Spectrum}} configuration.}

\mypar{Presto}
Like Starling, Presto is a SQL execution engine designed to execute on data on ``in situ'', on raw storage, rather than having users load data into specialized formats. We use a cluster of \texttt{r4.8xlarge} nodes to execute queries query. Each node has 32 threads and 244GB of DRAM. We used Amazon EMR~\cite{emr} 5.24.1 to set up Presto~\cite{presto} 0.219. Since we could have set this cluster up ourselves without using EMR we report only the cost of running EC2~\cite{ec2} virtual machines, and not the additional cost of Amazon EMR. We enable spilling to disk and query optimization. Before running queries, we collect statistics on all tables to provide data for the query optimizer.

We report performance and cost numbers on both a 5-node cluster (4 workers and 1 master) and a 17-node cluster, with 16 workers. We call these configurations \texttt{presto-4} and \texttt{presto-16} respectively.

\mypar{Amazon Athena}
Amazon Athena~\cite{athena} is a managed query service based on Presto Version 0.172~\cite{athena-release-notes}.  Users define a schema for data sitting in S3 buckets, then execute queries against this data without provisioning hardware. Users execute queries on Athena with a REST interface, when the query completes, results are written to an S3 object. For our results below, we report the run-time of each query as measured by the Athena Service. 

Users are charged for the number of bytes in S3 that their query scans. Like Starling, Athena has near zero cost when idle, with the exception of S3 costs for data storage. Unlike Starling, however, Athena provides users with no control over the degree of parallelism used in executing queries, and, as we will show, occasionally is unable to allocate sufficient resources to execute queries over large data files at all.

\mypar{Starling}
Unless otherwise noted, we configure Starling with all performance optimizations enabled including doublewrite (see Section~\ref{subsec:visibility_stragglers}) and pipelining (Section~\ref{subsec:pipelining}). Because Starling does not have a query optimizer, we set the number of tasks per stage manually as follows. First each input object is assigned a task. We then choose the number of tasks for join stages by sweeping over several options. We choose a configuration that is neither the fastest nor least expensive, but offers a good tradeoff of cost and performance. For the scale factor 1,000 dataset we disable multi-stage shuffling, but enable it for large joins in the scale factor 10,000 experiment. Starling's query plans use join orders generated from the Redshift optimizer with the \texttt{redshift-dc-dd} configuration, e.g. without partitioned data.
Unless indicated, configurations are fixed over experiments.

 \subsection{Cost of Operation}
\label{subsec:eval_cost_comparison}

\begin{figure}
  \begin{subfigure}[b]{\columnwidth}
    \centering
    \includegraphics[width=0.95\columnwidth, trim={0.25cm .25cm 0.25cm 0.25cm},clip]{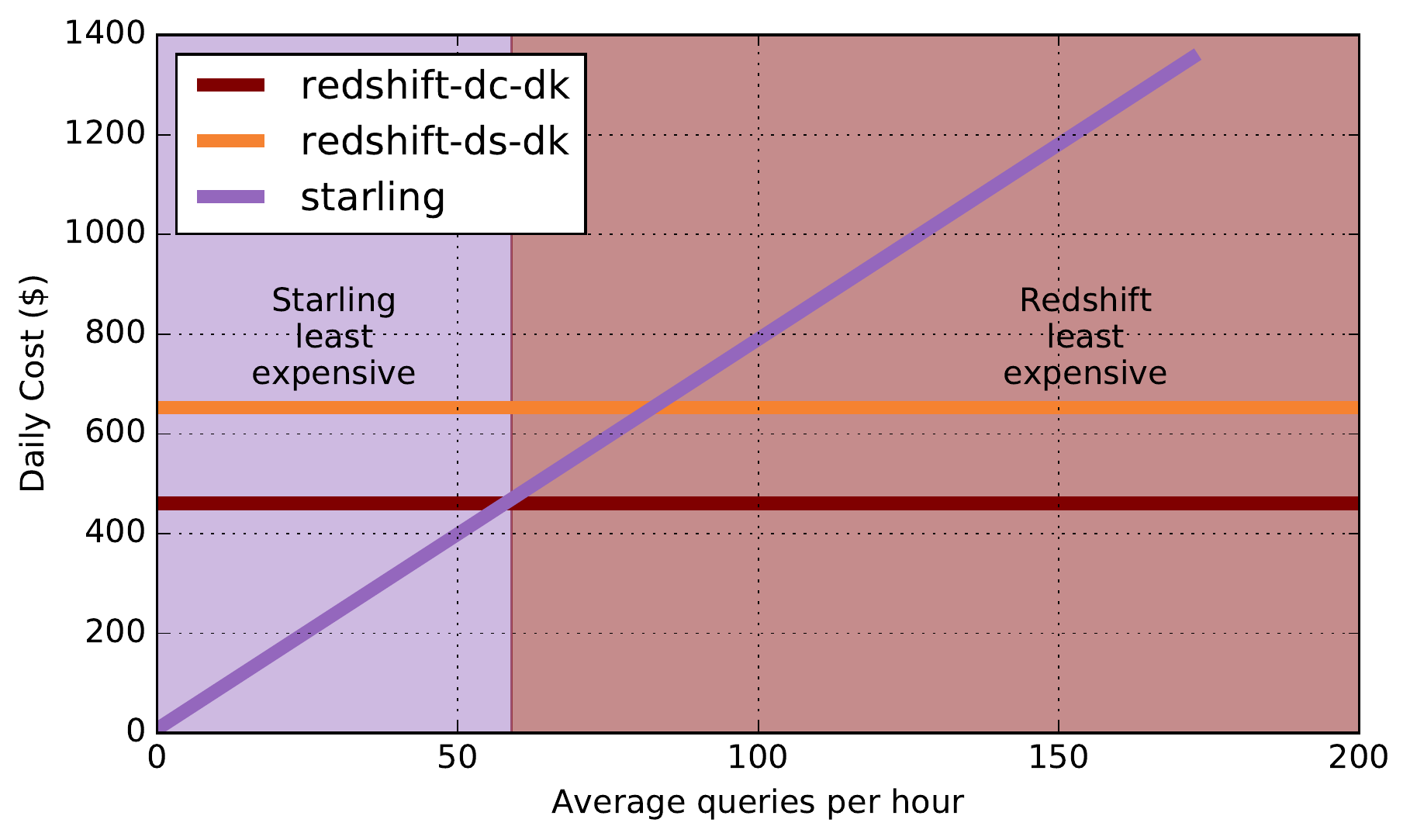}
    \subcaption{Starling vs configurations with data stored locally}
    \label{fig:cost_comparison_redshift}
  \end{subfigure}
  \begin{subfigure}[b]{\columnwidth}
    \centering
    \includegraphics[width=0.95\columnwidth, trim={0.25cm .25cm 0.25cm 0.25cm},clip]{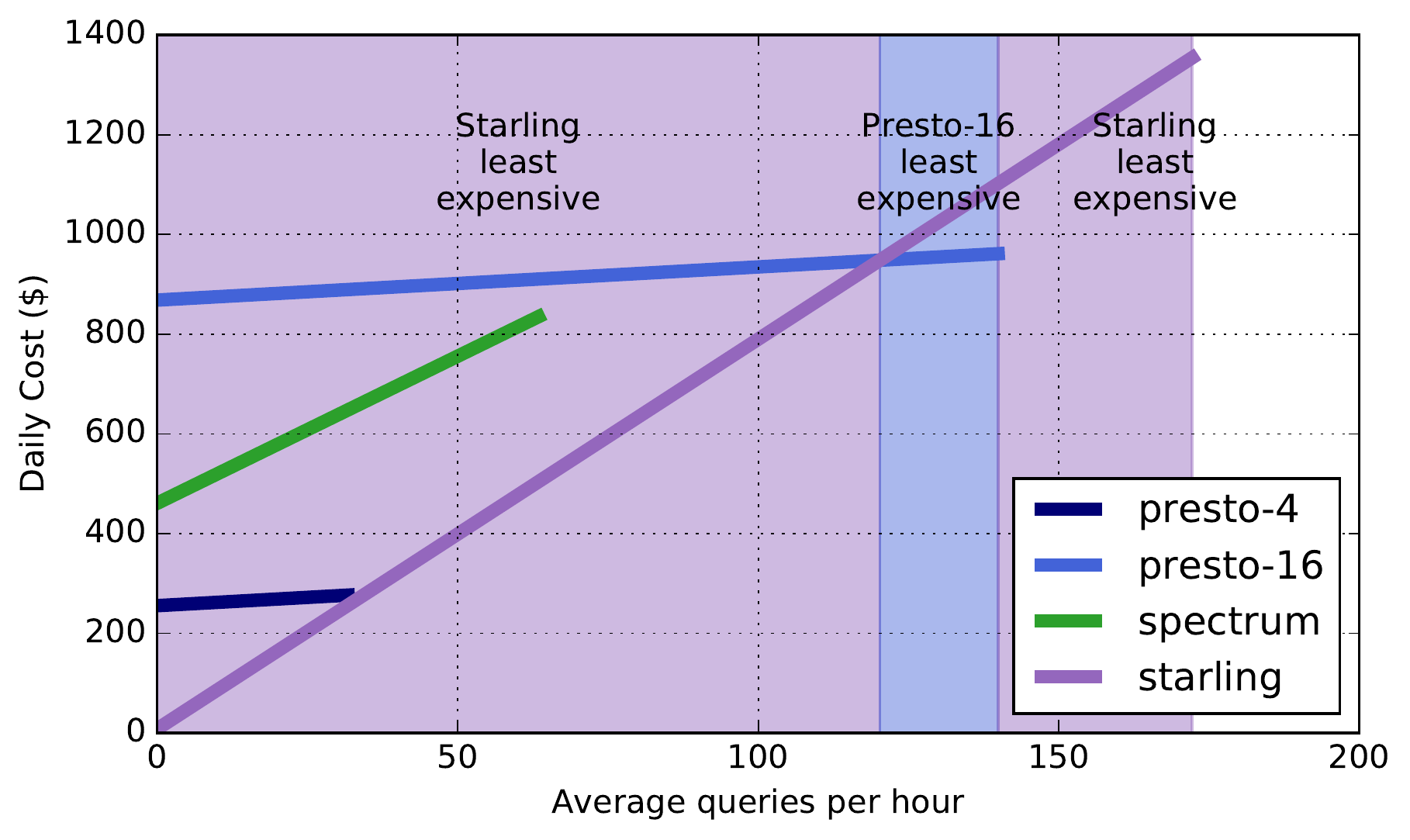}
    \subcaption{Starling vs configurations with data stored in S3}
    \label{fig:cost_comparison_s3}
  \end{subfigure}
  \caption{Daily cost of Starling and alternatives}
\end{figure}

\begin{figure*}
  \centering
  \includegraphics[width=0.95\textwidth, trim={0.25cm .25cm 0.25cm 0.25cm},clip]{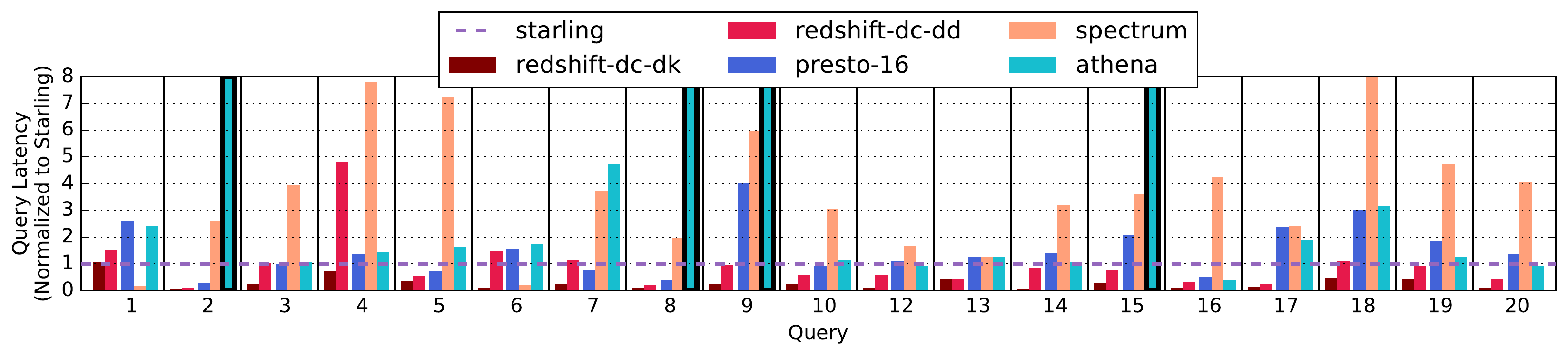}
  \caption{Query latency for several systems on the 1TB TPC-H dataset, normalized to Starling. Athena does not complete queries 2, 8 9, and 15, indicated by black outlines}
  \label{fig:query_latency}
\end{figure*}

We conduct experiments to demonstrate that at moderate query rates,
Starling is less expensive than alternatives.

We consider the daily cost of running a provisioned cluster with data stored locally compared to the cost of Starling as we change the number of queries executed per hour. We use the geometric mean of queries in the workload to determine a typical query execution time. Figure~\ref{fig:cost_comparison_redshift} compares the daily cost of running Starling to Redshift configurations with data pre-loaded and stored locally. The length of the line indicates the maximum number of queries that each system can execute back to back. Because the default distribution schema has lower performance than with a distribution key, but has the same daily cost, we leave it off the plot for simplicity. Starling incurs very little cost when queries are not executing but cost increases as the number of queries increase. \texttt{redshift-dc-dk} extends off right of the figure to 774, and \texttt{redshift-ds-dk} extends to 330. All Redshift configurations with data stored locally have fixed cost no matter how many queries are executed, hence there is a point where running Redshift becomes less expensive than running Starling, at around 60 queries per hour on the 1TB dataset.  Because Redshift has loaded and indexed
the data, its most efficient configuration is able to run queries faster than Starling, as we will show in Section~\ref{subsec:eval_cost_performance}.  Note that in addition to potentially incurring significantly higher operation costs, especially for infrequent queries, this performance requires pre-loading data and carefully tuning the database.

\begin{figure}
  \centering
  \includegraphics[width=0.95\columnwidth, trim={0 0.25cm 0 0},clip]{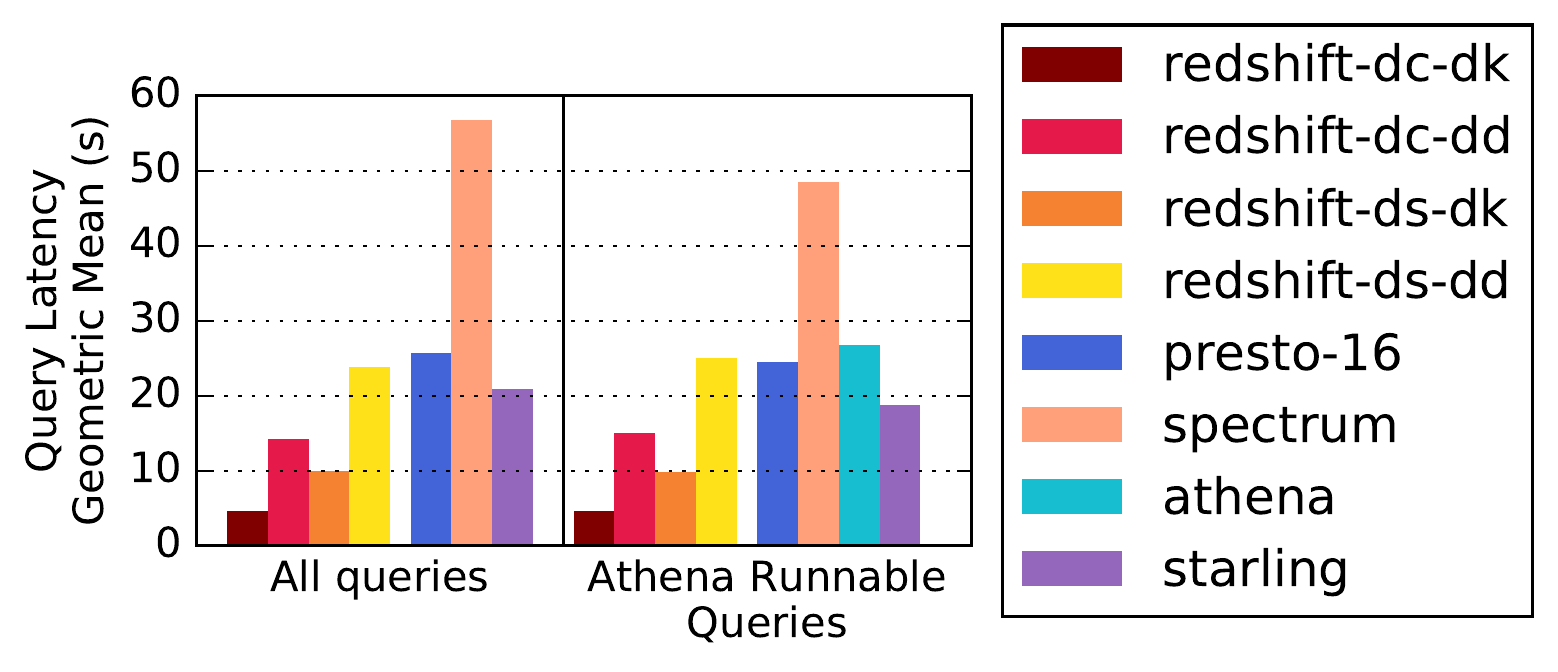}
  \caption{Geometric mean of latency on 1TB dataset}
  \label{fig:query_latency_gmean}
\end{figure}

We next compare against systems that, like Starling, read data directly from
cloud object storage. Figure~\ref{fig:cost_comparison_s3} compares two Presto
clusters of differing size, Redshift Spectrum, reading from S3, and Starling. Each
of these systems incurs more cost as more queries are executed. In the case of
Starling this cost comes from reading base table and intermediate data from S3,
as well as AWS Lambda costs. At about 120 queries per hour Starling's cost
increases beyond \texttt{presto-16}. However, as \texttt{presto-16} is not as
performant as Starling, with more than 153 queries per hour Starling is the only
system able to keep up. No configuration reading base tables from S3 can execute
more than 189 queries per hour back-to-back. While \texttt{presto-4} is cost
competitive with Starling at about 33 queries per hour, this is its maximum
query throughput when executing queries back-to-back. We discuss query latency
in more detail in the following section.

\begin{figure}
  \centering
  \includegraphics[width=0.95\columnwidth, trim={0 0.25cm 0 0.25},clip]{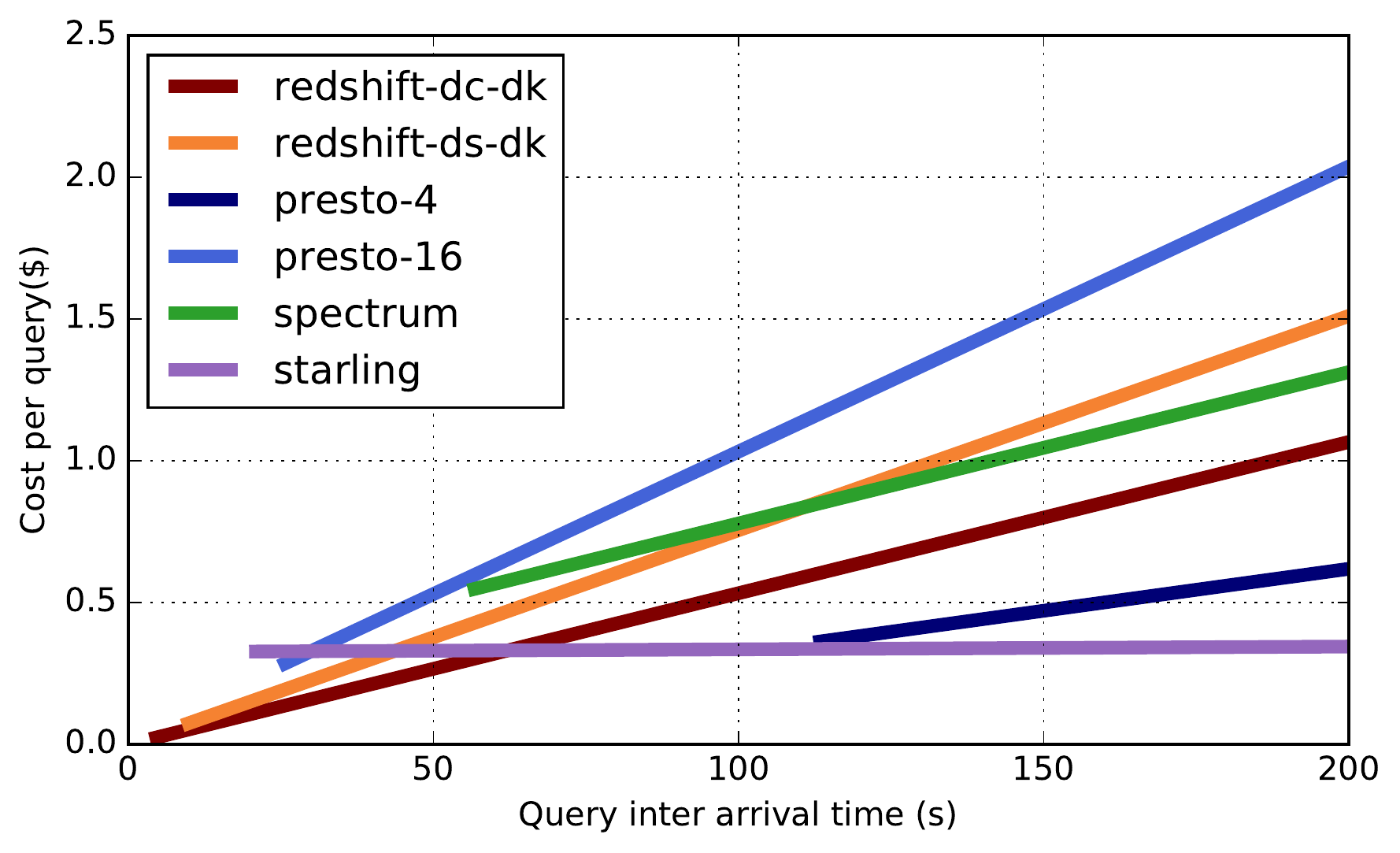}
  \caption{Cost per query (geometric mean of executed queries) on 1TB dataset as inter arrival times increase. Lowest time represents sending queries back-to-back}
  \label{fig:cost_per_query}
\end{figure}

\noindent\textbf{Cost per Query: }
Figure~\ref{fig:cost_per_query}
compares systems on the cost-per-query as query inter-arrival time changes. 
Starling has a fixed cost-per-query as its only operational costs are the small
coordinator, \$8 per day and S3 storage costs.  Provisioned systems
cost-per-query increases rapidly as the time in between queries increases. 

In summary, Starling is the least expensive system of all configurations when query volumes
are moderate.

\subsection{Query Latency}
\label{subsec:eval_cost_performance}

While low cost is a benefit, users with ad-hoc query workloads also want
interactive performance. Figure~\ref{fig:query_latency_gmean} compares the
geometric mean of the latency of queries in the workload. \updateL{R5R9}{Athena did not
complete four of the queries, 2, 8, 9, and 15 reporting either a resource exhausted error, or because Athena was missing some SQL functionality} Therefore we also include a
separate plot of the geometric mean of the subset of queries Athena completed. A
full comparison of Starling to Athena follows in
Section~\ref{subsec:payg_systems}. These four queries are slower than average and
thus removing them decreases the geometric mean for most systems. While Starling
is just over four times slower than \texttt{redshift-dc-dk}, it does not have
the advantage of pre-partitioned base tables nor sorted data. When compared to a
configuration without this advantage \texttt{redshift-dc-dd} on the same cluster without pre-sorting
or pre-partitioning, Starling's latency is less than 50\% slower than Redshift after loading data into its native format.
\updateL{R5D6}{Starling does not preclude these optimizations. On pre-partitioned data Starling achieves a speedup of nearly 2x on Q12 over unpartitioned base tables.}
Figure~\ref{fig:query_latency}
contains a query-by-query comparison. 
We leave off \texttt{presto-4} as it has
significantly higher latency than \texttt{presto-16}. For queries that do simple
scans and aggregations like Q1 and Q6, Spectrum, which uses stateless workers like 
Starling for base table scans, achieves very low latency, even compared to
Redshift configurations with local data. For queries with expensive joins like Q9
Starling has latency similar to Redshift configurations with default distribution (configurations with \texttt{dd}). This is likely because of extra cost of shuffling data when base tables are not partitioned on the join key. 

For users that require very low query latency and are cost insensitive, a
provisioned system with pre-loaded local data and tuned schema is still the best
choice. But for ad-hoc analytics on data in cloud object storage, Starling has
the lowest query latency. Against systems with tables pre-loaded, sorted, and
stored locally, it has the lowest cost for query inter arrival times more than
60 seconds, and against systems with data stored in S3, Starling is less
expensive for times more than 30 seconds.

\subsection{Scalability}
\label{subsec:scalability}

\begin{figure}
  \centering
  \includegraphics[width=0.95\columnwidth, trim={0 0.25cm 0 0.25},clip]{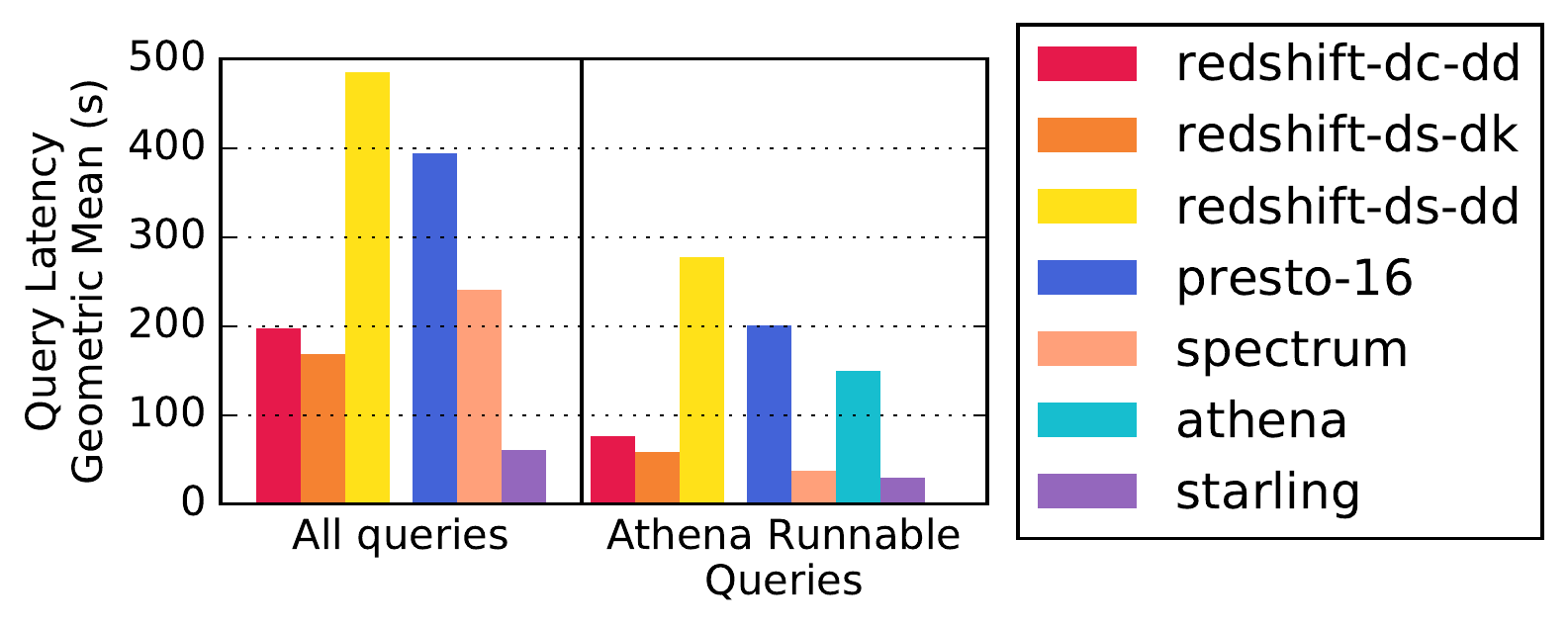}
  \caption{Geometric mean of latency on 10TB dataset}
  \label{fig:query_latency_big_gmean}
\end{figure}

In this section we show that Starling scales better than provisioned systems,
\update{R5D4}{even though it does not require expensive reprovisioning step that is required
to achieve good performance in other systems.}
We generated a scale factor 10,000 TPC-H dataset(10 TB before
compression) and executed a twelve queries from our query set, chosen for their
range in the size of input data and the number of joins . To scale to the 10TB
dataset, Starling increases the number of workers performing large joins, and
uses multi-stage shuffles to mitigate large S3 read costs.  We kept the
configuration of other systems the same. We summarize the results in
Figure~\ref{fig:query_latency_big_gmean}. In this case, each provisioned system
has query latency at least 2.7 times larger
than Starling.  Starling has the lowest latency of any compared system in 8 of
12 queries. \update{R5D4}{The next fastest configuration,
\texttt{redshift-ds-dk} has lower latency than Starling's for 1 of the 12
queries.} \texttt{redshift-dc-dk}, the best performing config on the 1TB
dataset, runs out of disk space when loading the 10TB dataset because of the
additional indices it builds during the data loading stage. Thus, we could not
compare against \texttt{redshift-\-dc-dk} for this experiment. 
Of course, adding more resources to the provisioned systems for this larger dataset
could allow them to execute with
lower latency. \update{R5D4}{This experiment demonstrates the challenges of provisioning a
system for an ad-hoc query workload with arbitrary amounts of data. For workloads
where the size of data inputs varies widely or is unknown ahead of time, the rapid 
elasticity of Starling gives it an advantage over provisioned systems.} While provisioned
systems would have to provision additional resources to handle more load,
Starling scales on a query-by-query basis and thus is able to be more flexible
to changes in input data size.

\begin{figure}[t]
  \centering
  \includegraphics[width=0.95\columnwidth, trim={0 0.25cm 0 0.25cm},clip]{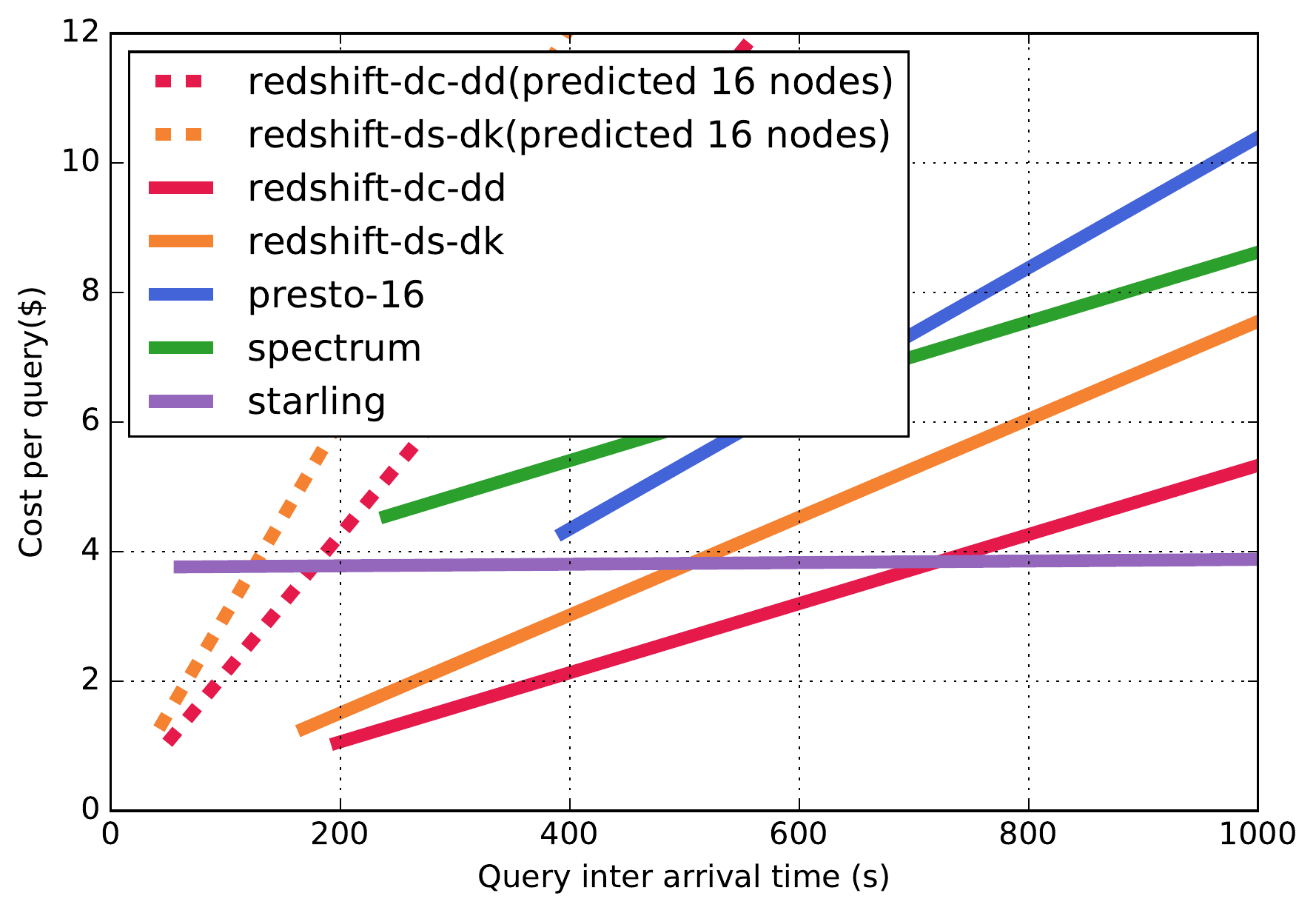}
\caption{Cost per query (geometric mean of executed queries) on 10TB TPC-H dataset}
  \label{fig:cost_per_query_big}
\end{figure}

Figure~\ref{fig:cost_per_query_big} compares the cost per query (which is fixed for
Starling) as query inter arrival time changes. The figure shows that Starling
has the lowest cost per query at any achievable query rate among all systems
that read data directly from S3. Even when comparing
against Redshift, which pre-loads data, Starling is less expensive when query
inter arrival times are 721 seconds apart or more, and achieves much higher
performance, partly because we did not scale up our Redshift cluster
when we scaled the data set. 

To understand what the cost per query would be had we scaled up Redshift, 
we assume that it would scale linearly with the number of nodes.  Under such an
assumption, with a 16-node 
cluster, its per query latency would similar to Starling, however, the dollar cost would
be 4x larger.  The 
dashed lines labeled \texttt{redshift-dc-dd (predicted 16 nodes)} and
\texttt{redshift-ds-dk (predicted 16 nodes)} in the figure show the estimated
cost-per-query of this configuration. Because the 16-node configuration's
cost-per-query is much higher, Starling is cheaper than Redshift ds-dk
when query interarrival times are 80 seconds or more.

Starling scales to larger datasets without requiring potentially
slow reconfiguration of provisioned systems.

\subsection{Concurrency}
\label{subsec:concurrency}

\begin{figure}
  \centering
  \includegraphics[width=0.95\columnwidth, trim={0.25cm .25cm 0.25cm 0.25cm},clip]{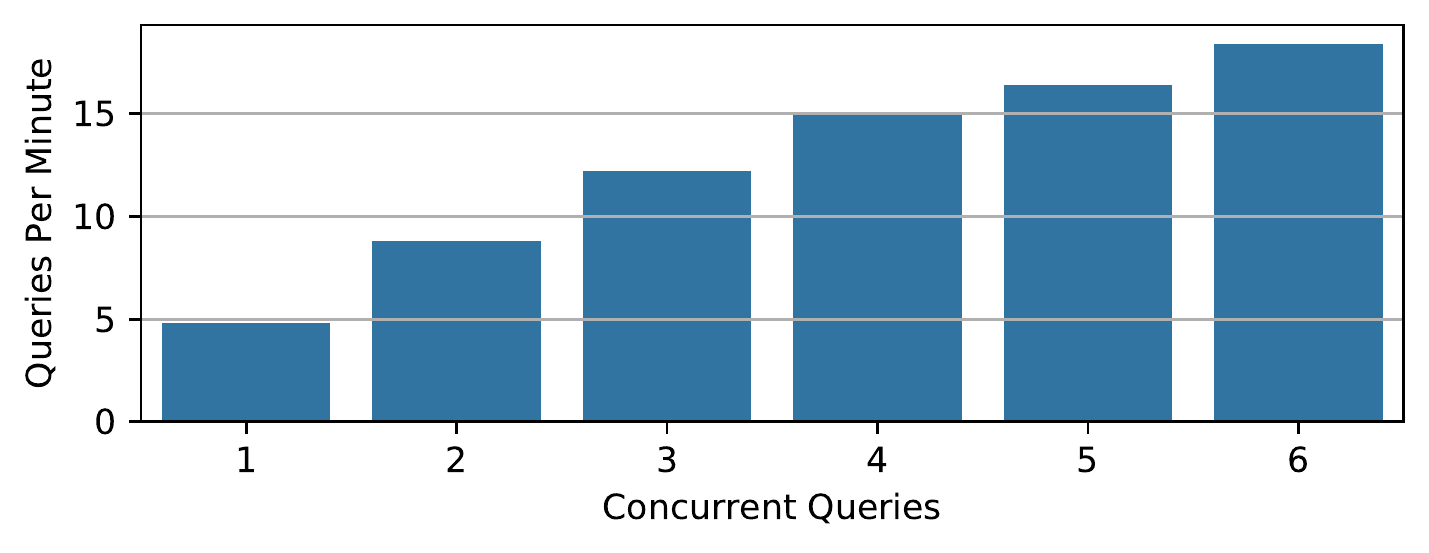}
  \caption{Starling Q12 Concurrency on TPC-H 1TB}
  \label{fig:concurrency}
\end{figure}

\updateL{R5D5, R9D2}{Starling is best suited for workloads with low to moderate query volumes. However, in cases where users need to run a burst of queries at once, Starling can scale to multiple concurrent queries. We show Starling's ability to scale with concurrent users by executing the same query, Q12, by multiple users. We see the results as we increase the number of concurrent users in Figure{~\ref{fig:concurrency}}. The maximum throughput is limited for two reasons. First, cloud function services have a limit on the number of concurrent function invocations. As we approach this limit, throughput will level off. Second, to support many concurrent queries the coordinator must invoke more and more functions in parallel using HTTP requests, straining resources on our low-cost coordinator. As high concurrency is not the focus of this work, we leave optimization of the coordinator as future work.} 

\subsection{Pay-per-query Services}
\label{subsec:payg_systems}

Fully managed query services, like Amazon Athena~\cite{athena}, come closest to realizing the goals of Starling. Users simply submit queries and charged on a query-by-query basis. Unfortunately, Athena is not a panacea for ad-hoc query workloads. Of the 12 queries on the 1TB dataset experiment Athena could not complete two. As seen in Figure~\ref{fig:query_latency_gmean}, for the queries that were completed by Athena, query latency was more than 50\% longer than with Starling. 

Despite these challenges, Athena has cost per query competitive with Starling.
Excluding the queries Athena was unable to execute, Athena has slightly higher cost per query than Starling at
the highest query rates, \$0.287 compared to Starling's \$0.0256.  Because
Starling has a small cost for the coordinator and Athena does not, Athena is cheaper at query rates lower than one query
every 350 seconds.

Athena does not scale to larger datasets.
Figure~\ref{fig:query_latency_big_gmean} shows that Athena only completed 5 out
of 12 queries.  On those completed, query latency was 5 times higher than
Starling's. Finally, Athena costs more than twice as much per query.  For users
who need interactive performance, the inability to tune for performance in
Athena makes it a nonstarter.

\subsection{Tunable Performance}
\label{subsec:eval_scheduling}

\begin{figure}[t]
  \centering
  \includegraphics[width=0.95\columnwidth, trim={0.25cm .25cm 0.25cm 0.25cm},clip]{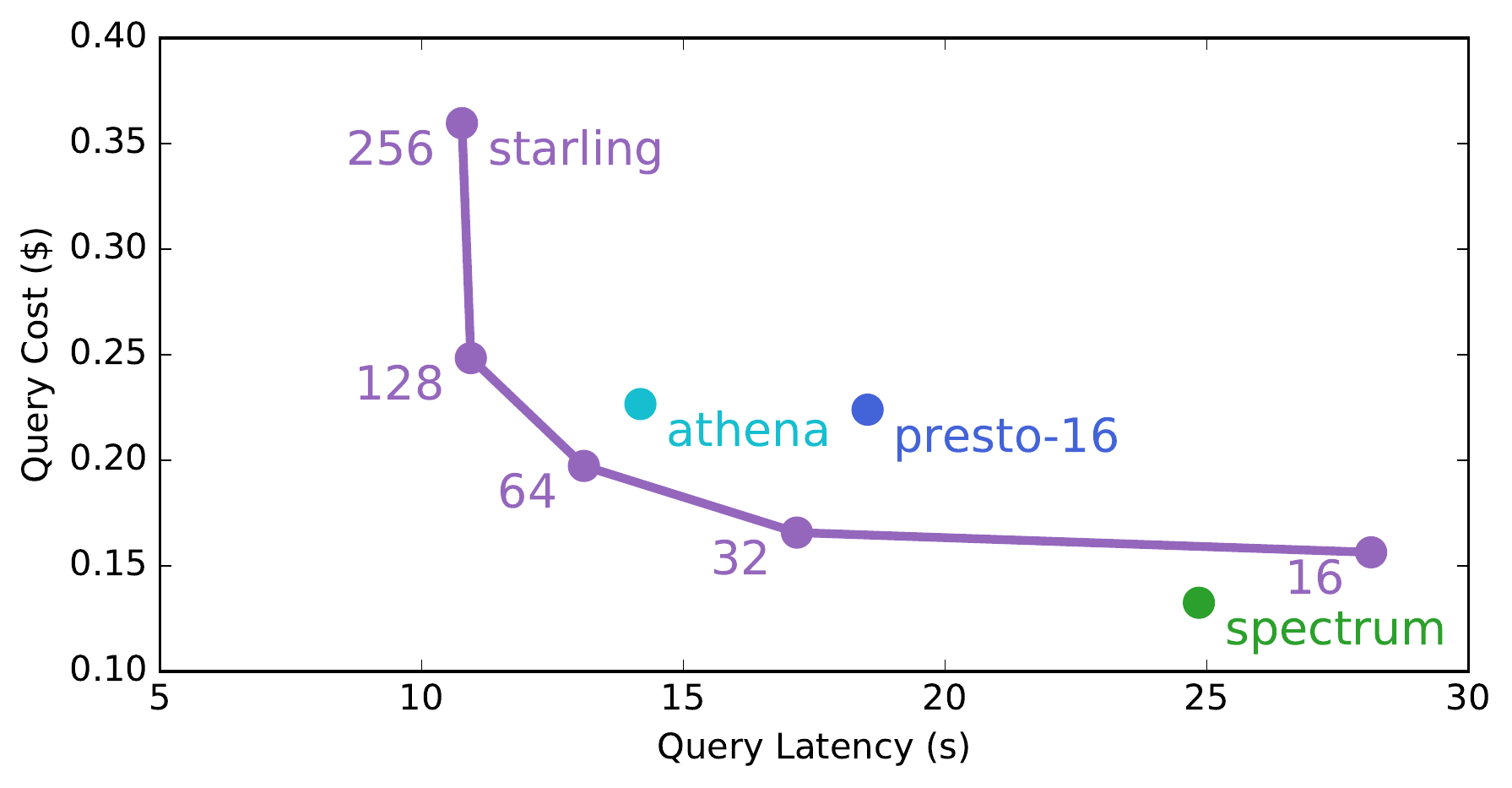}
  \caption{Cost and latency flexibility for TPC-H Q12}
  \label{fig:q12_cost_perf}
\end{figure}

Starling allows users to tune queries in order to reduce cost or increase
performance. We demonstrate this feature using TPC-H Query 12, a
select-project-join-aggregate query on the two largest tables in the dataset.
Figure~\ref{fig:q12_cost_perf} shows that Starling can increase performance for
additional cost. Each point represents the number of tasks used during the join
stage. The higher the
number of tasks, the higher the performance and cost. We connect the dots to
offer a visual aid. As we decrease the number of tasks, Starling is
dominated by the cost of execution time in AWS Lambda, but as we increase the
number of tasks, S3 read costs begin to dominate.

The cost of \texttt{presto-16} and \texttt{spectrum} represents the 
cost-per-query without idle time between queries. As we show in
Section~\ref{subsec:eval_cost_performance}, the cost per query increases with more idle time.

Even comparing with the strictest assumption of query arrival times Starling is
on the Pareto frontier for this query. It is less expensive than
\texttt{spectrum} and \texttt{athena} for the same performance. And it can
achieve higher performance than other systems reading from S3. 

Compared to provisioned systems, Starling is easier to tune for cost and
performance goals than other systems.

\subsection{Performance Optimizations}
\label{subsec:eval_perf_opt}

\begin{figure}[t]
  \centering
  \includegraphics[width=0.95\columnwidth, trim={0.25cm .5cm 0.25cm 0.25cm},clip]{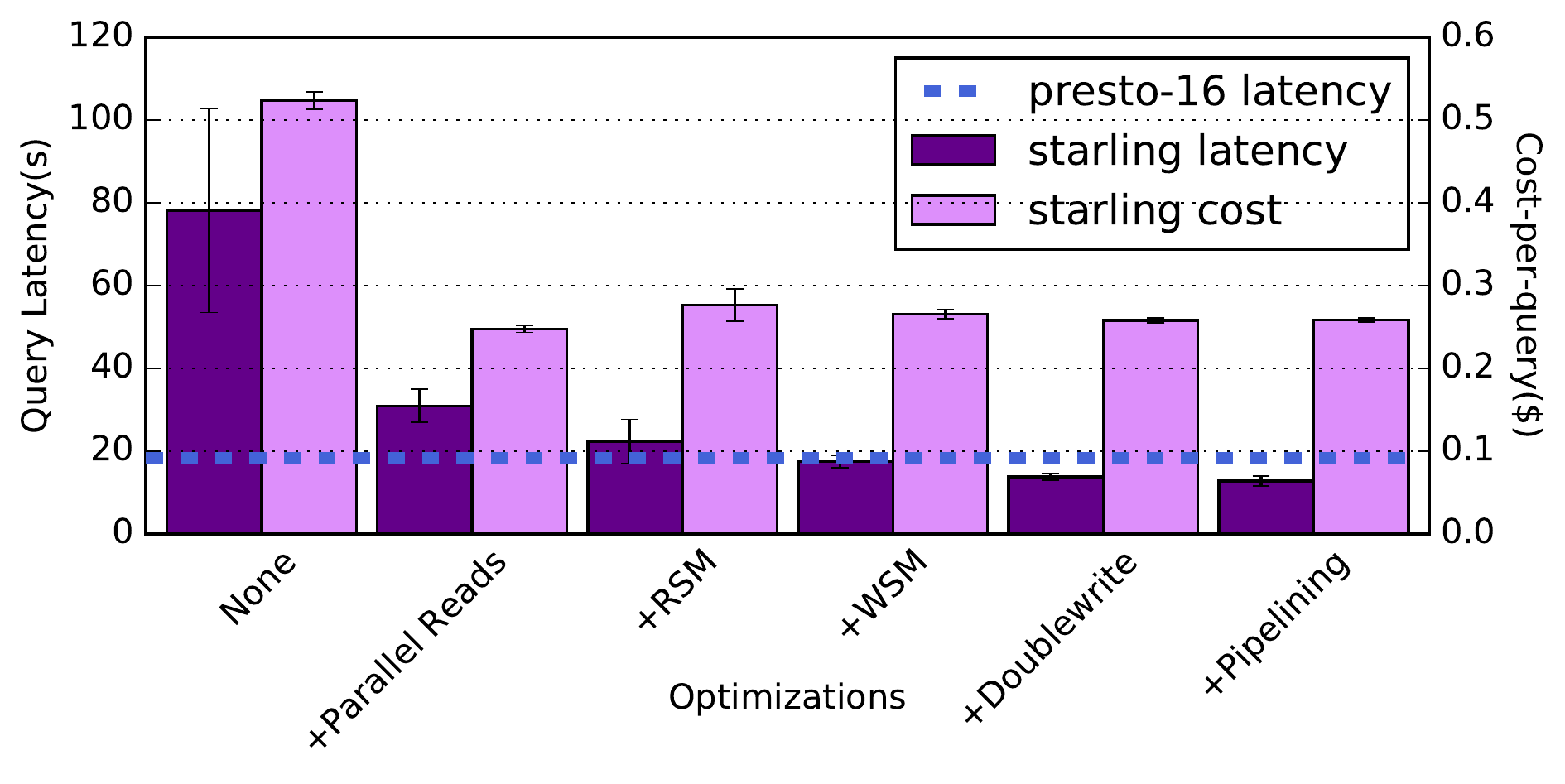}
\caption{Starling mean latency and cost on Q12 as optimizations are incrementally enabled. Error bars are standard deviation of ten executions}
\label{fig:optimizations}
\end{figure}

Without Starling's various performance optimizations, it cannot achieve
competitive latency with provisioned systems. In Figure~\ref{fig:optimizations},
we show Query 12's latency and cost as we add optimizations. 
For this query, we fix
the number of join workers to 128 and enable optimizations one-by-one as we move
right on the plot. Throughout the experiment, query cost remains approximately
constant. Query latency, however, decreases as expected. 
Without optimizations query execution time has very high
variance and takes about 80 seconds on average. Parallel reads, described in
Section~\ref{subsec:storage_latency}, have a big impact on performance,
particularly in the join stage as each worker makes hundreds of small reads.
Adding RSM and WSM, described in sections~\ref{subsec:read_straggler_mitigation}
and ~\ref{subsec:write_straggler_mitigation} respectively, decreases variance in
query runtimes and brings latency on par with \texttt{presto-16} by decreasing
variance in individual reads and writes. Finally, doublewrite, described in
Section~\ref{subsec:visibility_stragglers}, decreases the mean to just 12.8
seconds. This represents an improvement of six times over the mean runtime
without optimizations, more than 2.4 times with only parallel reads.

\section{Discussion: Starling's Cost to Cloud Providers}

In Section ~\ref{sec:evaluation} we demonstrated that for many workloads, Starling achieves the performance of a large system with significantly lower operational. However, the cost of each system is subject to the cloud provider's pricing strategy. For instance, since Athena is charged by the number of bytes read from S3, queries requiring extensive compute resources may be charged less than a trivial query that simply scans more data. Here we consider the case from the perspective of the cloud provider by making informed assumptions about the relative cost of resources. We find that with these assumptions, a system architected like Starling may be less resource intensive than traditional OLAP system architectures.

Amazon Athena~\cite{athena} is built on Presto~\cite{presto} version 0.172~\cite{athena-release-notes}. Although our \texttt{presto-16} configuration uses Presto 0.219, the latency of most queries on our presto cluster are close to \texttt{athena}. Because Athena uses limited statistics, we disabled statistics collection to compare query latency of Athena and \texttt{presto-16}. The geometric mean of all queries that executed on Athena on the 1TB dataset is within 3.5 seconds of \texttt{presto-16}, and all but 3 queries on Athena have latency within 4.25 seconds of \texttt{presto-16}. We attribute the remaining discrepancy to differences in hardware, configuration, and Presto version, but we do not have a way to determine Athena's internal configuration.

We established that a Presto cluster of 16 \texttt{r4.8xlarge} has performance similar to Athena. Thus, we assume that to execute an Athena query, the user has full use of a cluster of 16 nodes for the duration of the query. With this assumption we compare the computational cost of executing queries on an Athena-like system backed by Presto to Starling to determine which can achieve lower computation times, and thus lower cost, for the cloud provider. We compare Starling's execution time per core to \texttt{presto-16}. To be conservative, we assume that each AWS Lambda invocation has 2 vCPUs of compute. Each \texttt{r4.8xlarge} instance has 32 vCPUs. With 16 nodes, each second using the cluster consumes 512 core-seconds of compute.

\begin{figure}[t]
  \centering
  \includegraphics[width=0.95\columnwidth, trim={0.25cm .25cm 0.25cm 0.25cm},clip]{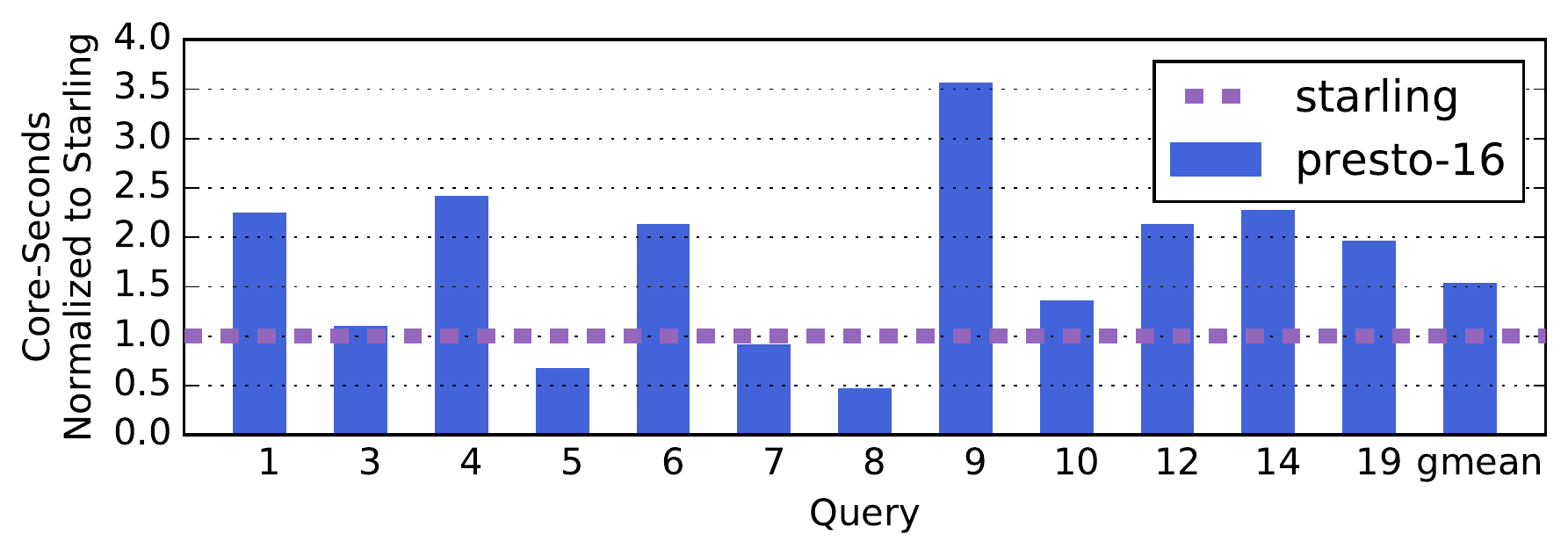}
  \caption{Total compute time in core-seconds}
  \label{fig:query_core_seconds}
\end{figure}

In Figure~\ref{fig:query_core_seconds} we compare the core-seconds Starling consumes per query to \texttt{presto-16}. Starling's join orders are taken from Redshift, have thus have the benefit of a query optimizer, so we compare the runtime of the optimized \texttt{presto-16} which has lower query latency than an unoptimized configuration. For most queries Starling consumes less compute for the same query compared to \texttt{presto-16}.

Although there are portions of a query that may be able to fully utilize a cluster of machines, some stages use only a subset of cores, or otherwise underutilize the cluster. In these cases, it may be more efficient for the cloud provider to provide a service that executes atop a general-purpose compute service like AWS Lambda rather than having to provision machines exclusively for query processing workloads that may end up underutilized.

Without detailed information on cluster architecture, load, and utilization, available only to cloud providers, it is difficult to say with certainty that Starling is able to achieve lower utilization for the cloud provider than a service like Athena. But using conservative estimates of CPU utilization we believe that this is likely. Therefore, it may be more efficient for cloud providers to offer query execution services with architectures similar to Starling in the future.

\section{Related Work}
\label{sec:related_work}

Starling is related to work in several different areas:

\noindent\textbf{Building stateful services on serverless platforms:} Cloud
function services have been used for implementing highly parallel
workloads~\cite{fouladi2017encoding, jonas2017occupy, gg}. PyWren~\cite{jonas2017occupy}, for instance,
  is a general-purpose tool for executing Python scripts in the cloud, but does not 
  demonstrate competitive performance nor cost for query processing workloads. GG~\cite{gg} is a framework for simplifying execution of jobs by managing tasks and stragglers on 
  cloud functions but is not specialized for query processing and thus can not make the
  same optimizations that Starling can with detailed knowledge of the workload.
ExCamera~\cite{fouladi2017encoding}
  is a tool for encoding video at low latency by exploding the high parallelism and low startup latency
  of cloud function services. Sprocket~\cite{ao2018sprocket} is a system for exploiting parallelism for
  video processing, but likewise has simple communication patterns. Unlike many of these
  workloads, query processing has more complex communication patterns, and can improve
  performance by making optimizations specific to query processing. In
particular, parallel query processing often requires shuffles to complete joins.
While some work has been done on shuffling data with function
services~\cite{pu2019shuffling}, it relies on having a set of provisioned
virtual machines to assist and does not execute full SQL queries.
Spark on Lambda~\cite{spark-on-lambda} was an attempt to run Spark jobs on AWS Lambda. However, this implementation was restricted to spark, does not implement straggler mitigation and does not demonstrate competitive performance.

\noindent\textbf{Cloud Analytical databases:} As customer move their analytical
workloads to the cloud, a myriad of systems have appeared to serve these
workloads. As we saw in Section 2~\ref{subsec:landscape}, none of the existing
systems fulfill the 3 key requirements of not pre-loading data, charge by 
query, and tunable performance. In particular, Starling is the first
analytical database engine built on top of a serverless platform.

\section{Conclusion}
\label{sec:conclusion}

In this paper we presented Starling, a query execution engine for data analytics built on cloud function services. Starling fills a gap in user's desire for an analytics system that provides interactive query latency at low cost per query for ad-hoc workloads while being tunable to performance and cost objectives. Starling achieves these goals by harnessing the rapid scaling and fine granularity that cloud function services provide. Starling overcomes the challenges of managing hundreds of stateless workers that must exchange data through opaque cloud storage. Starling's optimizations make it cost and performance competitive even with provisioned systems, and allow it to easily scale from small to large datasets and back without making explicit provisioning decisions. 

\balance

\bibliographystyle{abbrv}
\bibliography{lambda_analytics}

\end{document}